\documentclass[letterpaper]{aa520} 
\usepackage{natbib,amssymb,graphicx}
\bibpunct{(}{)}{;}{a}{}{,} 

\begin{document}

\title{Mass loss and rotational CO emission from Asymptotic Giant
  Branch stars}

\author{F. Kemper\inst{1,2}\thanks{SIRTF Fellow} \and R. Stark\inst{3} \and K.
  Justtanont\inst{4} \and A.~de~Koter\inst{1} \and
  A.G.G.M.~Tielens\inst{5,6} \and L.B.F.M.~Waters\inst{1,7} \and
  J.~Cami\inst{8} \and C.~Dijkstra \inst{1}}

\institute{Astronomical Institute ``Anton Pannekoek'', University of
  Amsterdam, Kruislaan 403, 1098 SJ Amsterdam, The Netherlands \and
  UCLA, Division of Astronomy, 405 Hilgard Avenue, Los Angeles, CA
  90095-1562, USA \and Max-Planck-Institut f\"ur Radioastronomie, Auf
  dem H\"ugel 69, D-53121 Bonn, Germany \and Stockholm Observatory,
  SCFAB, SE-106 91 Stockholm, Sweden \and Kapteijn Institute,
  University of Groningen, P.O. Box 800, 9700 AV Groningen, The
  Netherlands \and SRON Laboratory for Space Research, P.O. Box 800,
  9700 AV Groningen, The Netherlands \and Instituut voor Sterrenkunde,
  Katholieke Universiteit Leuven, Celestijnenlaan 200B, B-3001
  Heverlee, Belgium \and NASA Ames Research Center, Mail Stop 245-6,
  Moffett Field, CA 94035-1000, USA}

\offprints{F.~Kemper (kemper@astro.ucla.edu)}

\date{Received / Accepted}

\renewcommand{\theenumi}{\roman{enumi}}
\renewcommand{\labelenumi}{\bf\theenumi)}

\abstract{ We present submillimeter observations of rotational
  transitions of carbon monoxide from $J = 2 \rightarrow 1$ up to $7
  \rightarrow 6$ for a sample of Asymptotic Giant Branch stars and red
  supergiants.  It is the first time that the high transitions $J=6
  \rightarrow 5$ and $7 \rightarrow 6$ are included in such a study.
  With line radiative transfer calculations, we aim to determine the
  mass-loss history of these stars by fitting the CO line intensities.
  We find that the observed line intensities of the high transitions,
  including the $J=4 \rightarrow 3$ transition, are significantly lower
  than the predicted values. We conclude that the physical structure
  of the outflow of Asymptotic Giant Branch stars is more complex than
  previously thought. In order to understand the observed line
  intensities and profiles, a physical structure with a variable
  mass-loss rate and/or a gradient in stochastic gas velocity is
  required. A case study of the AGB star \object{WX Psc} is performed. We find
  that the CO line strengths may be explained by variations in
  mass-loss on time scales similar to those observed in the separated
  arc-like structures observed around post-AGB stars.  In addition, a
  gradient in the stochastic velocity may play a role. Until this has
  been sorted out fully, any mass loss determinations based upon
  single CO lines will remain suspect.}

\maketitle

\section{Introduction}
\label{sec:intro_co}

Low and intermediate mass stars ($1 < M < 8 \,M_{\odot}$) end their
life on the red giant branch and asymptotic giant branch \citep[AGB;
see][and references herein]{H_96_review}. During the AGB phase,
the stars have very extended tenuous atmospheres and shed almost their
entire hydrogen-rich envelope through a dense and dusty stellar wind.
In case of OH/IR stars, mass-loss rates can be so high that the dust
shell completely obscures the central star, and the object is
observable only at infrared wavelengths and through molecular line
emission at radio wavelengths.  The AGB phase is one of the few
occasions in stellar evolution when time scales are not driven by
nuclear (shell) burning but by surface mass loss. Helped by the low
surface gravity and strong stellar pulsations, gas can move away from
the star and will gradually cool. When the temperature drops below
$\sim$1400 K, dust formation occurs, and a dust driven wind will
develop. The mass-loss rates increase from $\dot{M} \approx 10^{-7}$
to a few times $10^{-5}$ $M_{\odot}$ yr$^{-1}$, while the AGB star
evolves from the Mira phase to an OH/IR star \citep{VH_88_IRAScolors}.
Recently, it has been suggested that higher mass-loss rates can be
achieved for oxygen-rich AGB stars. \citet{JST_96_OH26} find that OH
26.5+0.6 has undergone a recent increase in mass loss, leading to a
current rate of $5.5 \cdot 10^{-4} \, M_{\odot}$ yr$^{-1}$, a result
recently confirmed by \citet{FJM_02_oh26}.  Even higher mass-loss
rates were found for another oxygen-rich AGB star, \mbox{IRAS
  16342$-$3814}, for which the mass-loss rate may be as high as
$\sim$10$^{-3}$ $M_{\odot}$ yr$^{-1}$ \citep{DWK_02_IRAS16342}. A
similar rate of a few times 10$^{-3}$ $M_{\odot}$ yr$^{-1}$ is found
for the carbon-rich evolved star \object{AFGL~2688}
\citep{SMB_97_Egg}.

AGB stars are important contributors of dust to the interstellar
medium (ISM); it is estimated that a substantial fraction of the
interstellar dust is produced by oxygen-rich AGB stars
\citep[e.g.][]{G_89_stardust}.  In the outflow of evolved stars with an
oxygen-rich chemistry the dust composition is dominated by silicates,
both amorphous and crystalline \citep[e.g.][]{SKB_99_ohir,MWT_02_xsilI}.
The appearance of crystalline silicate features in the far-infrared
spectra of AGB stars seems to be correlated with a high optical depth
in the amorphous silicate resonance at 9.7 $\mu$m and hence a high
mass-loss rate \citep{WMJ_96_mineralogy,CJJ_98_ohir,SKB_99_ohir}.
This could be interpreted as evidence that a certain threshold value
for the density is required to form crystalline silicates. However,
\citet{KWD_01_xsilvsmdot} showed that observational selection effects
may play an important role in detecting crystalline silicates in AGB
stars with low mass-loss rates.  Therefore, the relation between
mass-loss rate and crystallinity remains unclear at present.

In order to further study the correlation between the wind density and
the dust composition, reliable mass-loss rates should be determined.
Mass-loss rates of AGB stars can be obtained from the thermal emission
from dust, predominantly coming from the warm inner regions
\citep[e.g.][]{B_87_dustshells}. They can also be inferred from
observations of molecular transitions, in particular from CO
\citep[e.g.][]{KM_85_massloss}. A catalogue compiled by
\citet{LFO_93_CO} lists observations of the CO $J=1 \rightarrow 0$ and
$J = 2 \rightarrow 1$ transitions of both O-rich and C-rich AGB stars.
(Hereafter we will use for these rotational transitions the notation
CO(1$-$0) etc.)  The mass-loss rates of a large number of objects from
the catalogue are derived. However, the derived mass-loss rates seem
to be underestimated for OH/IR stars, compared to the dust mass loss.
\citet{HFO_90_deficiency} have studied the correlation between IRAS
colours and mass-loss rates derived from CO(2$-$1) and CO(1$-$0)
observations.  In the case of very massive dust shells, they find that
the intensity of the CO(1$-$0) transition is too low compared to the
CO(2$-$1) transition, which they suspect to be due to a mass-loss rate
increase over time. This then hints towards a superwind phase, which
is generally believed to be important in the evolution of a Mira
towards an OH/IR star \citep[e.g.][and references
herein]{IR_83_AGBevolution}.  The superwind model was initially
introduced to explain the amount of mass seen in planetary nebulae
assuming that Miras are the progenitors of these nebulae
\citep{R_81_superwind}. Miras are believed to evolve into OH/IR stars
when they suddenly increase their mass-loss rate with a factor of
$\sim$100.

As the inner regions are warmer they are better probed by higher
rotational transitions. Thus a sudden density jump should be detectable
in the CO lines. Model calculations by \citet{JST_96_OH26} have
demonstrated this effect for \object{OH 26.5+0.6}, using
observations of rotational transitions up to  CO(4$-$3).
Unfortunately this transition is not sufficiently
high to firmly establish the recent onset  of a superwind, as
its excitation temperature is only 55 K. Nevertheless, \citet{JST_96_OH26} 
found that the peak
intensities of these lines were significantly higher than what could
be expected based on the extrapolation of the observed line strength
of the CO(2$-$1) transition and the upper limit obtained
for the CO(1$-$0) transition, assuming a constant mass-loss rate.
Similar results are reported for other AGB stars
\citep[e.g.][]{G_94_OH32OH44,DKF_97_superwind}. 

The work presented here aims to determine the mass-loss history of a
number of oxygen-rich AGB stars with an intermediate or high optical
depth in the near- and mid-infrared. For the first time, observations
of rotational transitions up to CO (7$-$6) have been
obtained ($T_{\mathrm{ex}} = 155$ K) which probe the more recent
mass-loss phases. In Sect.~\ref{sec:obs} we describe the observations
and data analysis. Sect.~\ref{sec:conditions} describes the model.
Our results are discussed in Sect.~\ref{sec:analysis}. Concluding
remarks and an outlook to future work is presented in
Sect.~\ref{sec:disc}.

\section{Observations and data reduction}
\label{sec:obs}

\subsection{Instrumental set-up}
\label{sec:setup}

\begin{table}
\caption{Technical details of the JCMT heterodyne receivers. The columns
list the used receivers, the frequency windows at which they operate, the
observable CO rotational transition, the beam efficiency $\eta_{\mathrm{mb}}$ and the half power beam width (HPBW).}
\begin{center}
\begin{tabular}{l c c c c}
\hline
\hline
receiver & Frequency & CO transition & $\eta_{\mathrm{mb}}$ & HPBW \\
         & (GHz)     &               &                      & \\
\hline
A3       & 215--275  & CO(2$-$1)     & 0.69                 & $19.7''$ \\
B3       & 315--373  & CO(3$-$2)     & 0.63                 & $13.2''$ \\ 
W/C      & 430--510  & CO(4$-$3)     & 0.52                 & $10.8''$ \\
W/D      & 630--710  & CO(6$-$5)     & 0.30                 & $8.0''$ \\
E        & 790--840  & CO(7$-$6)     & 0.24                 & $6.0''$ \\
\hline
\hline
\end{tabular}
\end{center}
\label{tab:efficiencies}
\end{table}

Observations of the $^{12}$CO(2$-$1), (3$-$2), (4$-$3), (6$-$5) and
(7$-$6) rotational transitions in the outflow of evolved stars were
obtained during several observing periods between April 2000 and
September 2002 using the
\emph{James Clerk Maxwell Telescope} (JCMT) on Mauna Kea, Hawaii. For
this purpose, all five different heterodyne receivers available at the
JCMT were used, including the new MPIfR/SRON E-band receiver which
operates in the \mbox{790--840 GHz} frequency range. A description of
this new receiver is given in Sect.~\ref{sec:E-band}. The technical
details and beam properties of the JCMT set up with the appropriate
heterodyne receivers are summarized in Table~\ref{tab:efficiencies}.
Observations with the B3- and W-receivers were performed in double
sideband (DSB) and dual polarization mode.  The DSB mode was also used
for the observations with the MPIfR/SRON E-band receiver. The
bandwidth configuration of the receiver, and hence the spectral
resolution was determined by the expected line width of the CO lines.
We used bandwidths of at least twice the expected line width to have a
sufficiently broad region for baseline subtraction.  Estimates for the
line width -- which is determined by the outflow velocity -- were
based on published values of line widths of the CO(1$-$0) transition
\citep[e.g.][and references herein]{LFO_93_CO}.

We used the beam-switching technique to eliminate the background. The
secondary mirror was chopped in azimuthal direction over an angle of
120$''$. Over these small angles the noise from the sky is assumed to be
constant. In case of extended sources we used a beam-switch of 180$''$.

\subsection{The MPIfR/SRON 800 GHz receiver}
\label{sec:E-band}

The observations of the CO(7$-$6) line were made with the MPIfR/SRON
\mbox{800 GHz} receiver in October 2001. This PI system is in
operation at the JCMT Cassegrain focus cabin since spring 2000.  The
receiver consists of a single-channel fixed-tuned waveguide mixer with
a diagonal horn. The mixer consists of a Nb SIS junction with NbTiN
and Al wiring layers fabricated at the University of Groningen, The
Netherlands.  Details on the fabrication of similar devices can be
found in \citet{JDL_00_E}.  Measured receiver temperatures at the
cryostat window are \mbox{$T_{\rm Rx}\simeq 550$ K} DSB. The receiver
has an intermediate frequency of $2.5-4$ GHz. System temperatures
including atmospheric losses varied between 6000--14000 K (SSB) at the
time of the observations.  The beam shape and efficiency have been
determined through observations of Mars and yield a deconvolved half
power beam width (HPBW) of 6$''$ and a main beam efficiency
$\eta_{\mathrm{mb}}$ of 24\%.

\subsection{Observations and data reduction}
\label{sec:subobs}

Our sample of evolved stars is given in Table~\ref{tab:obslist}, which
also indicates the distances towards the programme stars. The sample
includes AGB stars and red supergiants. In Table~\ref{tab:obsdetails} an
overview of the observed transitions is given, including cumulative
integration times and the observing date.  The data were obtained over
a long period from April 2000 until September 2002 in flexible
observing mode, and are part of a larger ongoing 
programme. During the observations, spectra of CO spectral standards
used at the JCMT were also obtained.  If necessary, a multiplication
factor was applied to the observations of our sample stars, to correct
for variations in the atmospheric conditions.  These factors are
listed in Col.~4 of Table~\ref{tab:obsdetails} and are based on
measured standard spectra.  Reliable standards are only available for
the transitions observed with the A3-, B3- and W/C-receivers, for
which the flux calibration accuracy is around 10\%. For the W/D- and
MPIfR/SRON E-band reliable standards for our lines of interest are
lacking.  Therefore we estimate that the absolute flux calibration in
these bands has an accuracy of 30\%.

Table~\ref{tab:efficiencies} lists the beam efficiencies
$\eta_{\mathrm{mb}}$ for all receivers.  The main beam
temperatures were calculated according to $T_{\mathrm{mb}} =
T_{\mathrm{A}}^{\ast} / \eta_{\mathrm{mb}}$, where
$T_{\mathrm{A}}^{\ast}$ is the measured antenna temperature.  These
main beam temperatures can directly be compared to observations from
other telescopes.

\begin{figure}
  \centerline{\includegraphics[width=8.5cm]{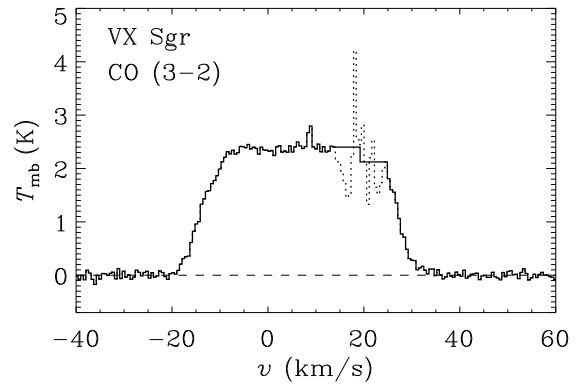}}
  \caption{Correction of the profile of the CO(3$-$2) transition 
    of \object{VX Sgr}. The dotted line represents the observation in which the
    interstellar contribution is clearly visible. Ignoring the
    interstellar contribution results in the solid line, which is used
    to obtain the integrated intensity.}
  \label{fig:correct}
\end{figure}

The reduced data is presented in Table~\ref{tab:transitions}.  A linear
baseline has been subtracted from the raw data, and the spectrum has
been rebinned to improve the signal-to-noise ratio. We aimed to cover
the line profile with at least $\sim$80 bins, which limits the
rebinning factor. The bin sizes after rebinning and the corresponding
r.m.s.~values are listed in Cols.~4 and 5 of
Table~\ref{tab:transitions}. Emission lines were detected in almost all
observations, except for \object{$\alpha$ Sco} CO(3$-$2) and \object{OH~104.9+2.4}
CO(6$-$5) and (7$-$6), for which we only obtained upper limits on the
main beam temperatures. The line profiles of all transitions are shown
in Figs.~\ref{fig:wxpsc}--\ref{fig:oh104}.  In some cases,
interstellar lines are visible in the spectrum, for example in
\object{VX Sgr}. To determine the integrated intensities we have cut
the interstellar lines out of the spectrum, and interpolated both
parts of the spectrum, as is demonstrated in Fig.~\ref{fig:correct}.
The resulting profile was integrated to obtain $I$, which is the
integrated intensity in K km s$^{-1}$.  The system velocity
$V_{\mathrm{LSR}}$ and the terminal expansion velocity $v_{\infty}$
are estimated directly from the line profile.  The lines show a wide
variety of shapes. There are parabolic line profiles, like those of
\object{WX~Psc} (Fig.~\ref{fig:wxpsc}), \object{IRC+50137}
(Fig.~\ref{fig:irc50}), \object{AFGL~5379} (Fig.~\ref{fig:gl}),
\object{CRL~2199} (Fig.~\ref{fig:crl}) and \object{OH~104.9+2.4}
(Fig.~\ref{fig:oh104}). These parabolic line profiles indicate that
the lines are optically thick \citep{M_80_CO}. On the other hand, many
objects show signs the double-horned profiles indicative of an
optically thin molecular layer. The most illustrative example is
\object{VY~CMa}, in which the CO(2$-$1) and CO(3$-$2) transitions
clearly show a double-peaked profile, although the peak around the
central velocity indicates a more complex outflow structure
(Fig.~\ref{fig:vycma}). In addition, some flat-topped profiles are
observed, most notably those of \object{VX~Sgr}
(Fig.~\ref{fig:vxsgr}). These flat-topped line profiles are considered
to be characteristic of molecular layers which have $\tau \sim 1$ at
these frequencies \citep{M_80_CO}.

\citet{JST_96_OH26} have observed \object{OH 26.5+0.6} with the JCMT
as well and report that they find line intensities $I = 25.8$ and 36.0
K km s$^{-1}$ for the CO(3$-$2) and (4$-$3) transition respectively.
In addition they have scaled IRAM observations of the CO(2$-$1) to a
15m dish, to mimic the JCMT. The intensity of this line turned out to
be 7.8 K km s$^{-1}$.  Their results agree well with our results in
case of the CO(2$-$1) and (3$-$2) transition, but they have observed
an intensity of a factor of $\sim$2 higher for the CO(4$-$3)
transition. The origin of the discrepancy with our results is unknown.

\section{Physical conditions in the outflow: a model}
\label{sec:conditions}

The observed line profiles provide information on the physical
structure of the outflow of these AGB stars, as the spectral
resolution at the observed frequencies is sufficiently high to resolve
the velocity structure.  The terminal expansion velocity $v_{\infty}$
can be derived directly from the width of the line profile
(Table~\ref{tab:transitions}). The model we use to analyze the CO data
is based on a study by \citet{S_88_molspec} and was previously used by
\citet{JCT_94_redgiantwinds}. The interpretation of our observations
using this model is discussed in Sect.~\ref{sec:analysis}.

\subsection{Description of the model}
\label{sec:modelco}

The code consists of two parts: The first part solves the radiation
transfer equation in the co-moving frame \citep{MKH_75_twolevel},
computes the level populations (in full non-LTE)
and iterates until level populations and radiation field are
consistent.  For solving the level populations, a Newton-Raphson
method is used \citep{SH_86_multilevel}.  The calculations take into
account (de-)excitation through collisions, of which the rate is
defined by the thermal velocity distribution, calculated from the
local temperature, as well as (de-)excitation induced by a local
radiation field and spontaneous de-excitation.  The code treats pure
rotational transitions in the ground and first vibrational levels,
which are connected through these collisional and radiative
transitions. The model can calculate the populations of as many as 50
levels at once, and is also applicable to molecules other
than CO.  The non-LTE rate equations to determine the level
populations are described by

\begin{eqnarray}
& & n_i \sum_{j \neq i} (A_{ij} + C_{ij} + B_{ij} \overline{J}_{ij}) - \nonumber\\
& & \qquad \qquad \sum_{j \neq i} n_j (A_{ji} + C_{ji} + B_{ji} \overline{J}_{ij}) = 0
\label{eq:rateq}
\end{eqnarray}

A change from level $i$ to level $j$ can be induced by collisional
transitions (with the collisional rate $C_{ij}$) and radiative
transitions, including spontaneous emission ($A_{ij}$, where $A_{ij} =
0$ for $i < j$) and stimulated emission and absorption
($B_{ij}\overline{J}_{ij}$). The collisional transition rates $C_{ij}$
are taken from laboratory measurements and potentials calculations
\citep{FL_85_ratecoefficients} and are extended up to $J = 30$.

The line profile integrated mean intensity $\overline{J}_{ij}$
consists of two components:

\begin{enumerate}
  
\item The \emph{continuum} radiation, originating from dust locally
  present. This radiation field can be switched off, by assuming there
  is no dust present in the considered part of the outflow.
 
\item Line radiation originating from a local region. The size of
  this region is defined by a velocity which \citet{S_88_molspec} and
  also \citet{JCT_94_redgiantwinds} have referred to as
  \emph{stochastic} velocity $v_{\mathrm{sto}}$.  The nature of this
  \emph{stochastic} velocity is not specified, but
  physically should consist of a thermal component
  $v_{\mathrm{therm}}$ and a turbulent component $v_{\mathrm{turb}}$,
  given by

\begin{equation}
v_{\mathrm{sto}} = \sqrt{(v_{\mathrm{therm}})^{2}+ (v_{\mathrm{turb}})^{2}}
\label{eq:velocities}
\end{equation}

In the outflow, the stochastic velocity is assumed to be constant and
in almost all cases dominated by turbulence. The effect of the
stochastic velocity is Doppler broadening of the lines, which is taken
into account in the radiative transfer.

\end{enumerate}

In the second part of the code, the calculated level populations are
used as input to determine the observable line profiles by
ray-tracing. Again the \emph{stochastic} velocity is used, this time
to determine the width of the interaction region along the
line-of-sight to the observer.  Integration over the full beam, for
which the telescope parameters are required, yields the emergent
line profile.

\subsection{Free parameters}
\label{sec:parameters}

The model has a number of free parameters (see
Table~\ref{tab:standard}). In this section we will discuss the various
parameters and their relevance for the model calculations.

\subsubsection{Density profile}
  
The density profile $\rho(r)$ of the outflow determines the collision
probabilities and optical depths required to solve
Eqs.~(\ref{eq:rateq}) and to calculate the line profiles. The density
profiles follows from the equation of mass continuity
\begin{equation}
\rho(r) = \frac{\dot{M}}{4 \pi r^2 \, v_{\mathrm{exp}}(r)}
\label{eq:masscont}
\end{equation}
where the expansion (or outflow) velocity profile used in the model is
defined by
\begin{equation}
v_{\mathrm{exp}} (r) = v_{\infty} \bigg( 1 - \frac{b}{r} \bigg)
\label{eq:vel}
\end{equation}
In this equation $v_{\infty}$ represents the terminal velocity.
Constant $b$ is chosen such that the expansion velocity at the stellar
surface is given by $v_{\mathrm{exp}}(R_{\ast}) = 10^{-2} v_{\infty}$.
The density structure is set by the following input parameters

\begin{enumerate}
  
\item The gas \emph{mass-loss rate} $\dot{M}$ determines the mass
  input at the inner radius of the circumstellar shell. Our model
  allows us to simulate the effect of a time-variable mass-loss rate
  introducing \emph{one jump} in the mass-loss history at an arbitrary
  point in the outflow ($r_{\mathrm{superwind}}$), where the density
  can increase or decrease with a specified factor.  Except for this
  jump the mass-loss rate is constant, and therefore the density
  profile scales with the current mass-loss rate at \mbox{$r <
    r_{\mathrm{superwind}}$} and with the past mass-loss rate at $r >
  r_{\mathrm{superwind}}$.
  
\item The density profile also scales with the outflow
  velocity profile given in Eq.~(\ref{eq:vel}), which is fixed by
  the \emph{terminal velocity} $v_{\infty}$.
  
\item The \emph{stellar radius} $R_{\ast}$ determines the
  base of the wind. The density $\rho(R_{\ast})$ at the inner radius
  follows from $R_{\ast}$, $\dot{M}$ and $v_{\exp}(R_{\ast})$ using
  the equation of mass continuity (\ref{eq:masscont}).
  
\item The \emph{outer radius} $R_{\mathrm{out}}$ denotes the extent of
  the outflow.

\end{enumerate}

\subsubsection{Temperature profile}

The \emph{temperature profile} $T(r)$ is another important parameter
that influences the level populations in the circumstellar CO, by means of 
collisions. The
temperature profile may be compiled self-consistently, i.e.~based
on calculations of realistic heating and cooling processes
\citep[e.g.][]{GS_76_OHIR,JCT_94_redgiantwinds,CN_95_water,ZE_00_WHya}.
As a first order estimate we have used a power law of the form $T(r)
\propto r^{-\alpha}$, where the index
$\alpha$ depends on the mass-loss rate and is derived from the outer
regions of the temperature profiles calculated by
\citet{JCT_94_redgiantwinds}.
  
\subsubsection{Dust-to-gas ratio and dust properties}

Unfortunately it is difficult to study the gas and dust mass-loss rate
completely independent from each other, as continuum emission from
dust may have an effect on the \mbox{(de-)excitation} rates, as
described in Sect.~\ref{sec:modelco}.  In particular, infrared photons
at 4.6 $\mu$m pump CO molecules from the ground vibrational state
$v=0$ to the first vibrational level $v=1$
\citep[e.g.][]{M_80_CO,S_88_molspec}. The molecules will eventually
de-excite to the vibrational ground level, but not necessarily to the
same rotational ground level. This causes a higher population of the
higher CO rotational levels than which reflects the kinetic
temperature of the gas and the line radiation field.  As the source of
the 4.6 $\mu$m radiation is predominantly thermal dust emission, the
\emph{dust-to-gas ratio} and the \emph{dust opacity} are important
input parameters. For simplicity, we assumed that there was no dust
present in the outflows.  For some of the calculations we did include
dust to study the effect on the line strengths. In those cases we used
a dust opacity corresponding to the mixture of solid state components
derived for \object{OH 127.8+0.0}, a typical OH/IR star
\citep{KDW_02_composition}.  The same power-law temperature
distribution as for the gas is used to calculate the thermal emission
from the grains, although this is most likely not true.
  
\subsubsection{Velocity field}

The velocity field has already been mentioned as a constraint for the
density structure, but it also plays an important role in the
formation of line profiles.  The outflow velocity profile
(constrained by the \emph{terminal velocity} $v_{\infty}$ and the
velocity law given in Eq.~(\ref{eq:vel})) and the \emph{stochastic
  velocity} $v_{\mathrm{sto}}$ determine the location and extent of
the interaction regions. As said before, the stochastic velocity is
assumed to be constant throughout the dust shell.

\subsubsection{Distance and telescope parameters}
  
The resulting main beam temperatures depend on the \emph{distance}
towards the object. In addition, the telescope \emph{beam size} is
important to determine what part of the object falls inside the beam.
In case the circumstellar shell is resolved, the \emph{pointing
  displacement} (usually 0$''$) should be known as well.

\section{Analysis of the results}
\label{sec:analysis}

\begin{figure}
\sidecaption
  \includegraphics[width=8.5cm]{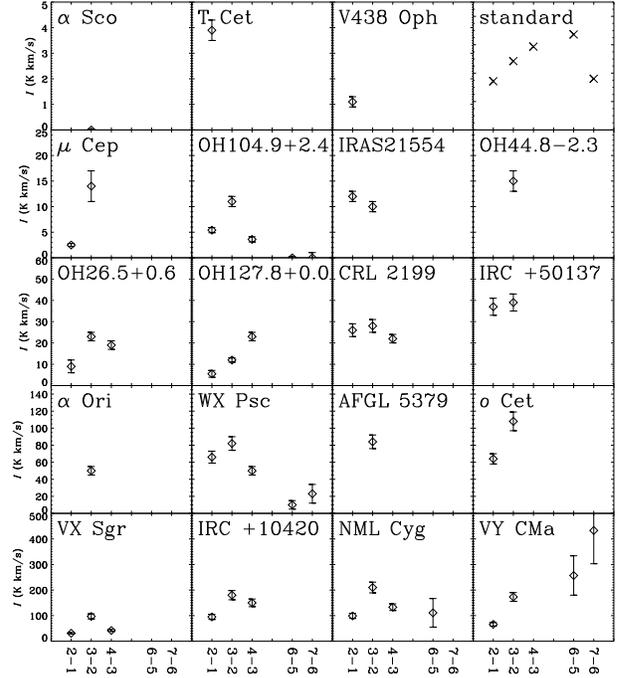}
  \caption{Overview of integrated intensities for each line observed in 
    our programme stars. The horizontal axis of each panel lists the
    rotational transitions observed, where the spacing between the
    tick marks is proportional to the difference in frequency. On the
    vertical axis the integrated intensity (K km s$^{-1}$) is given.
    The diamonds represent the measured values; in addition the error
    bars are shown (data from Table~\ref{tab:transitions}). Note that
    only the line strengths of \object{VY~CMa} increase with higher
    rotational transitions.  For most of the other stars (except
    \object{OH~127.8+0.0}, \object{IRC~+50137} and \object{IRAS
      21554+6204}) CO (3$-$2) is the brightest line. In the upper
    right corner the relative values for the standard model (see
    Table~\ref{tab:standard}, Fig.~\ref{fig:standard}) are presented
    for comparison (indicated with $\times$ symbols).}
  \label{fig:sed}
\end{figure}

\begin{figure}
  \centerline{\includegraphics[width=8.5cm]{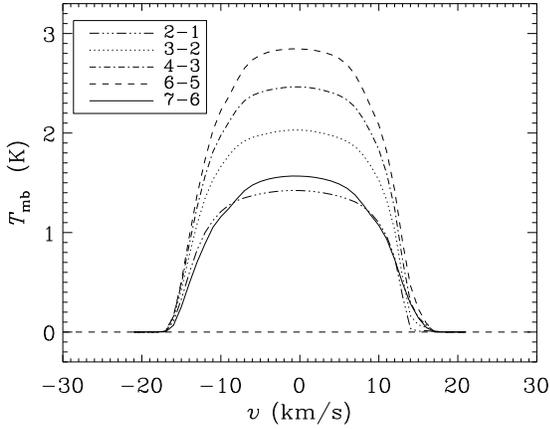}}
  \caption{Line profiles calculated for a standard AGB model, folded with the JCMT beams. 
    The model parameters are given in Table~\ref{tab:standard}. }
  \label{fig:standard}
\end{figure}

Here we will analyze the observations using the model described in
Sect.~\ref{sec:conditions}. Fig.~\ref{fig:sed} shows the intensities
integrated over line width of each observed line for all our sample
stars.  For all sources, except \object{VY~CMa}, the integrated
intensity increases from the CO(2$-$1) to (3$-$2) transition, and
decreases again for higher transitions.  This is also visible in the
peak main beam temperatures ${T_{\mathrm{mb}}}$
(Table~\ref{tab:transitions}).  Since most studies concentrate on the
lower transitions (up to CO(3$-$2)) this was not noticed before. An
exception are the JCMT observations of \object{OH~26.5+0.6} performed
by \citet{JST_96_OH26}, where the CO(4$-$3) transition is included as
well. However, as pointed out in Sect.~\ref{sec:subobs}, they observed
an increasing line strength with increasing rotational transition.
This differs from our observations of this object that the CO(3$-$2) is
the strongest emission line.

In order to explain the observational trends, we have constructed a
standard model assuming physical parameters widely used for AGB
outflows (see Table~\ref{tab:standard}). We used a mass-loss rate of
$\dot{M} = 10^{-5}$ $M_{\odot}$ yr$^{-1}$, and calculated the level
populations of the CO gas between the stellar radius $R_{\ast}= 4.0
\cdot 10^{13}$ cm and the outer radius \mbox{$R_{\mathrm{out}} = 6000
  \, R_{\ast}$}. For the terminal velocity we used $v_{\infty} = 15.0$
km s$^{-1}$ and the turbulent velocity was assumed to be
$v_{\mathrm{sto}} = 1.0$ km s$^{-1}$. A power-law temperature profile
was chosen: $T(r) = 2000 \, (r/R_{\ast})^{-0.7}$ K. We used for the
relative abundance of the CO gas with respect to molecular hydrogen
[CO]/[H$_2$] = $3.0\cdot 10^{-4}$, and we ignored the contribution of
thermal emission from dust to the local radiation field.  Finally, we
placed this system at a distance of 1000 pc, and used the JCMT
telescope parameters to calculate the emerging line profiles
(Table~\ref{tab:standard}, Fig.~\ref{fig:standard}). The lines show
increasing peak and integrated intensities with increasing line
strengths, up to CO(6$-$5). The CO(7$-$6) line is
again much weaker which can be explained by the relatively narrow HPBW
of the E-band (Table~\ref{tab:efficiencies}). This transition is
comparable in strength to the CO(2$-$1) transition, for this
standard set of parameters.  This is a general characteristic of all
other studies calculating the line intensities for commonly used AGB
parameters
\citep[e.g.][]{G_94_revisedmodel,G_94_OH32OH44,JCT_94_redgiantwinds}.

\begin{table}
\caption{Parameters of the standard AGB model}
\begin{center}
\begin{tabular}{l l}
\hline
\hline
parameter                      & value\\
\hline
distance                       & 1.0 kpc           \\
$v_{\infty}$                   & 15.0 km s$^{-1}$   \\
$v_{\mathrm{sto}}$             & 1.00 km s$^{-1}$ \\
$R_{\mathrm{in}}$              & 5 R$_{\ast}$     \\
$R_{\mathrm{out}}$             & 6000 R$_{\ast}$     \\
$R_{\ast}$                     & $4.0 \cdot 10^{13}$ cm    \\
$\dot{M} $                     & $10^{-5}$ $M_{\odot}$ yr$^{-1}$\\
$T(r)$                         & 2000 $(r/R_{\ast})^{-0.7}$ K \\
$[\mathrm{CO}]/[\mathrm{H}_2]$ & $3.0\cdot 10^{-4}$\\
dust-to-gas ratio              & 0\%\\
\hline
\hline
\end{tabular}
\label{tab:standard}
\end{center}
\end{table}

In the following sections we will try to find a set of parameters to
explain our observations: in general the CO(3$-$2) is the
strongest line, which contradicts the results of the standard model.
In order to study as many stars as possible in a systematic way, we
will use a line ratio diagram based on the CO(3$-$2)/CO(2$-$1) and
CO(4$-$3)/CO(2$-$1) ratios of integrated intensities, rather than
trying to fit the intensities and line profiles.

The two low mass-loss rate AGB stars \object{RV Boo} and \object{X
  Her} are added to the sample; these stars are the only ones for
which sufficient reliable line ratios of interest can be derived from
published JCMT data (see Table~\ref{tab:litvalues}).  For
\object{RV Boo} the ratios of the integrated intensities are 1.5 and
2.0 for CO(3$-$2)/CO(2$-$1) and CO(4$-$3)/CO(2$-$1) respectively
\citep{BKO_00_rvboo}. For \object{X Her} these numbers are 2.1 and 3.4
respectively \citep{KYL_98_COsurvey,KO_99_COcatalogue}. The
observations of \citet{KYL_98_COsurvey} were obtained with the
  CalTech Submillimeter Observatory (CSO) and are rescaled to 
  the JCMT observations of \citet{KO_99_COcatalogue} such that the line
intensities of the CO(3$-$2) transitions both reflect the same dish
size  and can be compared to our CO(3$-$2) and CO(2$-$1)
  observations. One should bear in mind however, that \object{RV
    Boo} and \object{X Her} are not representative of AGB stars with
  spherical outflows.  \citet{KJ_96_XHer} have mapped \object{X Her}
  in CO lines and conclude that in addition to a slow spherically
  expanding shell there are indications for bipolar outflows with a
  higher velocity, which carry a significant fraction of the ejected
  gas. This result is confirmed by \citet{KOP_03_massloss}, who in
  addition present SiO line observations indicative of a circumstellar
  rotating disk.  Interferometric CO line observations of \object{RV
    Boo} indicate that this object also has a disk, possibly showing
  Keplerian rotation \citep{BKO_00_rvboo}.  
Therefore, comparison of these stars with our data and analysis should be
  done with some reservation.

In the literature, we found a sample of six Miras, which were
  observed in all three lines discussed here, using CSO (see
  Table~\ref{tab:litvalues}). It is possible to scale these
  observations to the JCMT observations by accounting for the dish
  size.  However, we have chosen not to do this, because it is unknown
  how reliable the rescaled data still is, as little is known about
  the beam filling factor of the various transitions, while the beam
  sizes of the telescopes are very different. Instead we chose to
  compare these observations with our model calculations, as will be
  discussed in Sect.~\ref{sec:mdotco}.

\subsection{A constant mass-loss rate?}
\label{sec:mdotco}

\begin{figure}
  \centerline{\includegraphics[width=8cm]{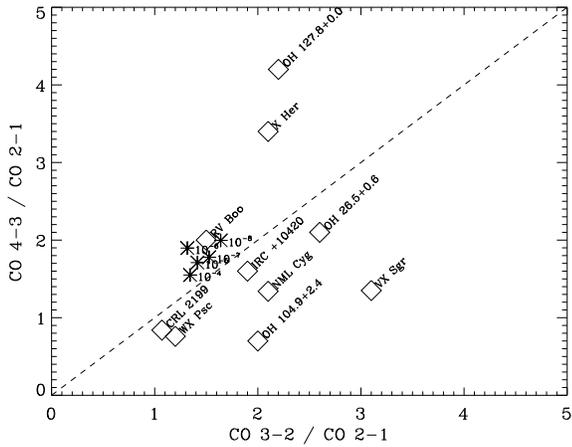}}
  \caption{Line ratio diagram. On the horizontal axis the ratio of the 
    integrated intensities of CO(3$-$2)/CO(2$-$1) is given, while the
    CO(4$-$3)/CO(2$-$1) ratio is plotted on the vertical axis. The
    diamonds represent the positions of our sample stars, complemented
    with literature data for \object{X Her} and \object{RV Boo}. Equal
    ratios are indicated with the dashed line. In case of our sample
    stars, the CO(3$-$2) transition is stronger than the (4$-$3)
    transition, therefore all observations can be found in the lower
    right half of the diagram. The only exception is \object{OH
      127.8+0.0} which is found in the upper left half.  The asterisks
    mark the positions of model calculations, where we used the
    standard parameters (see Table~\ref{tab:standard}).  Only the
    mass-loss rate was varied and is given in units of $M_{\odot}$
    yr$^{-1}$.}
 
  \label{fig:mcst}
\end{figure}

In a line-ratio diagram (Fig.~\ref{fig:mcst}), the observed values
occupy the lower right half of the diagram, corresponding to the
region where the CO(3$-$2) line is stronger than the CO(4$-$3) line.
The values corresponding to \object{OH 127.8+0.0} are an exception and
are found in the upper left half. This data point should be treated
with care though, as the detected lines suffer from interference with
interstellar absorption (Fig.~\ref{fig:oh127}) and therefore the line
intensities are not well known (see Table~\ref{tab:transitions}).  The
literature data of \object{RV~Boo} and \object{X~Her} are also located
in the upper left half of the diagram.

First, we assume that the mass-loss rate is constant. For five
different mass-loss rates (10$^{-8}$, 10$^{-7}$, 10$^{-6}$, 10$^{-5}$
and 10$^{-4}$ $M_{\odot}$ yr$^{-1}$) we have calculated the emerging
line profiles, thus covering the full range in $\dot{M}$ from Miras to
OH/IR stars \citep{B_87_dustshells,VH_88_IRAScolors}.  All other input
parameters were assumed to have the standard values given in
Table~\ref{tab:standard}. The predicted CO(3$-$2)/CO(2$-$1) and
CO(4$-$3)/CO(2$-$1) ratios of the integrated intensities were compared
to the observed ratios.  

The model calculations (marked with asterisks) are found in the upper
left half of the diagram where the CO(4$-$3) line is stronger than the
CO(3$-$2) line, and are therefore not consistent with the observed
line ratios.  All model line ratios are found in a narrow range to one
end of the region where the observations are found (see
Fig.~\ref{fig:mcst}).  Only the observations of \object{RV~Boo} match
the modelled line ratios, but this could be merely a coincidence
  as \object{RV~Boo} is not a typical AGB star.

\begin{figure}
  \centerline{\includegraphics[width=8cm]{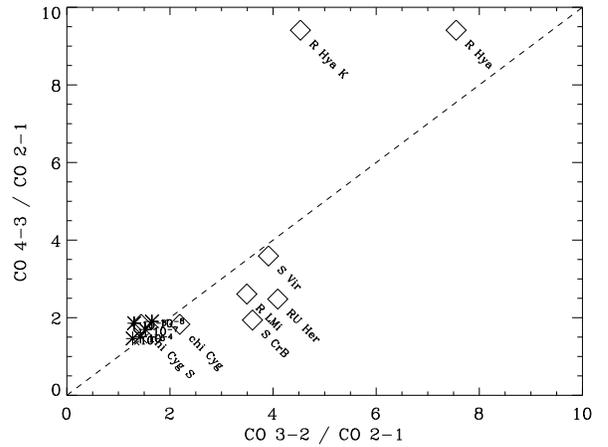}}
  \caption{Line ratio diagram for CSO data. On the horizontal axis the ratio of the 
      integrated intensities of CO(3$-$2)/CO(2$-$1) is given, while
      the CO(4$-$3)/CO(2$-$1) ratio is plotted on the vertical axis.
      The diamonds represent the ratios obtained for a sample of six
      Miras.  The CO(4$-$3) and CO(3$-$2) data are obtained from
      \citet{Y_95_CO3-2}, and the CO(2$-$1) data is taken from a study
      by \citet{KYL_98_COsurvey}. For \object{R Hya} and
      \object{$\chi$ Cyg} additional CSO CO(3$-$2) data are available,
      from \citet{KYL_98_COsurvey} and \citet{SKY_95_molecular}
      respectively. In the diagram these measurements are indicated
      with 'R Hya K' \citep{KYL_98_COsurvey} and an '$\chi$ Cyg S'
      \citep{SKY_95_molecular}.  Equal ratios are indicated with the
      dashed line. The asterisks mark the positions of model
      calculations, performed for the CSO beam and dish size, where we
      used the standard parameters (see Table~\ref{tab:standard}).
      Only the mass-loss rate was varied and is given in units of
      $M_{\odot}$ yr$^{-1}$.}
   \label{fig:cso}
\end{figure}

There are not many reports in the literature of AGB stars
  observed in these three lines with the JCMT, but we compared the
  results discussed here with observations performed using the CSO.
  For that purpose, we have recalculated the model line ratios for the
  CSO beam and dish size. A sample of six Miras is consistently
  observed with CSO, where the CO(4$-$3) and CO(3$-$2) measurements
  are obtained by \citet{Y_95_CO3-2}, and the CO(2$-$1) observations
  by \citet{KYL_98_COsurvey}. Additional CSO observations of the
  CO(3$-$2) line in two of these objects were also included
  \citep{KYL_98_COsurvey,SKY_95_molecular}. The results are shown in
  Fig.~\ref{fig:cso}. Similar to Fig.~\ref{fig:mcst}, the line ratios
  derived from the standard calculations are found just above the
  dashed line, indicating that the CO(4$-$3) line should be stronger
  than the CO(3$-$2) transition. However, most observations are found
  well below the dashed line, where the CO(3$-$2) is the strongest
  line.  \object{$\chi$ Cyg} falls in this region as well. However, if
  we use the measurement of \citet{SKY_95_molecular} for the CO(3$-$2)
  line, the line ratios become such that it is found in the same
  region as the model ratios. Possibly this point is unreliable, as it
  does not come from a consistent data set. \object{R Hya} seems to be
  an outlier for both CO(3$-$2) measurements. Another remarkable
  observation is that the observed CO(2$-$1) lines seem to be weaker
  than what is expected from the model calculations, given the fact
  that the calculated ratios are closer to the origin of the plot.  We
  may conclude that in general the CSO observations occupy more or
  less the same region of the plot with respect to the model ratios as
  our observations. Therefore, in the remainder of this paper we will
  limit our detailed analysis to our JCMT data.
 
Apparently, variations of the mass-loss rate alone do not
change the line ratios enough to significantly increase the strength
of the CO(3$-$2) line with respect to the CO(4$-$3) line.  In the next
section, we will investigate to what extent variations in the other
parameters can shift the model calculations such that the line
strength ratios more closely resemble the observed values.

\subsection{Exploring parameter space}
\label{sec:parspace}

To further explore parameter space, we opted to vary the input
parameters of the standard model (Table~\ref{tab:standard}) one by one,
and compare the line ratios with the observations. Combining the
changes in line ratios from variations in the individual parameters
then provides a feeling for the range in line ratios that can be covered,
and may show whether or not it is possible to explain the observed
line ratios at all. Of course, once a satisfactory match in line
ratios is achieved by combining the effects of changes in individual
parameters, fine tuning should be performed to fit the observed data
in detail.  This is necessary as some of these parameters might not be
completely independent from each other, and the precise combined
effect on the line profile is difficult to predict.

\begin{figure}
\sidecaption
  \includegraphics[width=8.5cm]{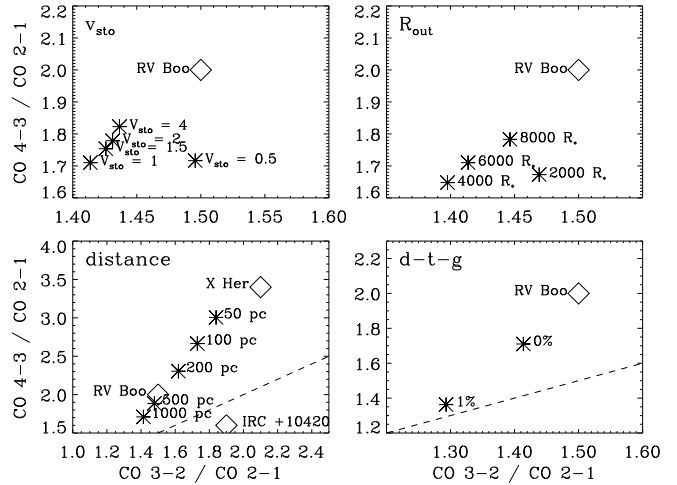}
  \caption{Mosaic of diagrams representing the 
    CO(4$-$3)/CO(2$-$1) ratio on the vertical axis versus the
    CO(3$-$2)/CO(2$-$1) ratio on the horizontal axis. We used the
    standard model described in Sect.~\ref{sec:analysis} and
    Table~\ref{tab:standard} and varied for each panel one of the
    parameters. From the upper left corner turning clockwise the
    investigated parameters are: the stochastic velocity
    $v_{\mathrm{sto}}$ (km s$^{-1}$), the outer radius
    $R_{\mathrm{out}}$ ($R_{\ast}$), the dust-to-gas ratio and the
    distance $D$ (pc). The line ratios resulting from the model
    calculations are marked with asterisks and the observed line
    ratios with diamonds. The dashed lines indicate equal line ratios.
    Note that the ranges plotted on the axes are smaller than the
    ranges in Fig.~\ref{fig:mcst} to improve readability.}
  \label{fig:mosaic1}
\end{figure}

In Fig.~\ref{fig:mosaic1} a mosaic of line-ratio diagrams is shown, in
which the effects of changes in the stochastic velocity, the outer
radius, the dust-to-gas ratio and the distance are shown. In general
the effects due to changes in these parameters are small. To keep the
plots readable, only small parts of the original line-ratio diagram
(Fig.~\ref{fig:mcst}) are shown. The modelled line ratios for which
these parameters are varied scatter mainly closely around the observed
values for \object{RV Boo}. In all these modelled line strengths, the
CO(3$-$2) line is still weaker than the CO(4$-$3) line. Of course
varying the parameters mentioned here causes changes in the absolute
line strengths, but the line ratios are not so much affected.

In the models where the distance was varied, we placed the object
progressively closer  to the observer, such that the beam filling factor
is initially less than unity, but increases with decreasing distance.
Although the beam size corresponding to the CO(2$-$1) transition is
larger than that corresponding to the CO(3$-$2) transition, the line
formation region of the CO(2$-$1) transition is located so much
further out that the object is first resolved for the CO(2$-$1)
transition. This implies that less emission from this line is received
by the telescope. When this happens, the line ratios increase. The
CO(4$-$3)/CO(2$-$1) line ratio increases faster for decreasing distance
than does CO(3$-$2)/CO(2$-$1), because the CO(3$-$2) line
emission is the next to become resolved, as this line is formed more
inwards in the circumstellar shell, but still further out than the
higher transitions.

The stochastic or turbulent velocity determines the interaction length
along the line-of-sight, i.e.~the region over which the line is formed
(see also Sect.~\ref{sec:modelco} and~\ref{sec:parameters}). The
effect of a larger turbulent velocity is different for optically thick
and optically thin lines. In the optically thin case, a change in
profile strength may result from changes in the line source function
in the (near and far) parts of the line interaction region, that is
added relative to the default case. In the optically thick case the
relevant source function is the one at the location where $\tau
\approx 1$, which shifts towards the observer when $v_{\mathrm{sto}}$
is increased. It may therefore differ from the default case. As these
effects tend in the same direction for all lines (except possibly when
lines change from optically thin to optically thick), the line ratios
are found not to change dramatically when varying the turbulent
velocity.

Changing the outer radius has a stronger effect on the line ratios, as
can be seen in Fig.~\ref{fig:mosaic1}. When the outer radius is
increased, more
relatively cold gas will be present. In this gas mostly the lower
rotational levels are populated, thus increasing predominantly the
CO(2$-$1) transition. The higher the transition, the less it is
affected by the outer radius.

The last parameter shown in Fig.~\ref{fig:mosaic1} is the dust-to-gas
ratio.  The most important effect of adding dust to the circumstellar
shell is in the population of the rotational levels. Continuum
emission at 4.6 $\mu$m can be absorbed by CO molecules, exciting them
from the ground to the first vibrational level.  They will return to
the vibrational ground state by spontaneous emission, but
preferentially to a higher rotational level than they started from.
This has a non-LTE effect on the level populations, leading to
variations in both the line strengths and the line ratios.

\begin{figure}
\sidecaption
\includegraphics[width=8.5cm]{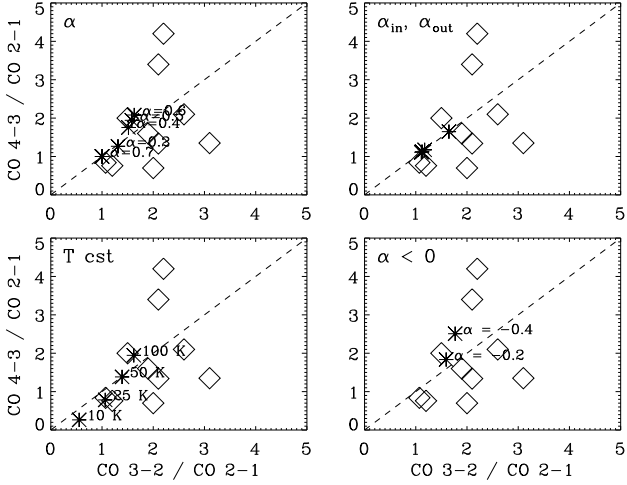}
  \caption{Mosaic of diagrams representing the 
    CO(4$-$3)/CO(2$-$1) ratio versus the CO(3$-$2)/CO(2$-$1) ratio.
    Again, the standard model parameters (Table~\ref{tab:standard})
    were used. Only the temperature profile was varied. From the panel
    in the upper left corner turning clockwise the adopted temperature
    profiles are: {\bf i)} $T(r) \propto r^{-\alpha}$, with $\alpha$
    indicated in the plot. {\bf ii)} $T(r)$ is described with a
    function consisting of two power-laws. See text for description.
    {\bf iii)} $T(r) \propto r^{-\alpha}$, with negative values of
    $\alpha$, indicated in the plot. {\bf iv)} A constant temperature
    throughout the circumstellar shell, where the adopted values are
    indicated in the plot.  Again, the predicted line ratios are
    marked with asterisks and the observed line ratios with diamonds.
    The diamonds are not labelled to avoid a crowded plot, but can
    easily be identified using Fig.~\ref{fig:mcst}. The dashed lines
    indicate equal line ratios.}
  \label{fig:mosaic2}
\end{figure}

We also investigated the effect of the temperature distribution. The
results are shown in Fig.~\ref{fig:mosaic2}. Various temperature
profiles have been used in the different panels of this figure.  The
simplest approach is to consider a power law $T(r) = T_{0}
(r/R_{\ast})^{-\alpha}$, where $\alpha$ is usually positive and has a
value around \mbox{0.5--0.6} for realistic profiles
\citep{JCT_94_redgiantwinds}.  We considered a much broader range of
$\alpha$, including negative values and also a constant temperature,
\mbox{i.e.~$\alpha=0$}. These cases are of course not a true physical
representation of the dust shell, but are just considered to study the
effect of extreme conditions.  In most cases, we used a temperature at
the inner edge of $T_{0} = 2000$ K.  However, when the power law is
shallow (small $\alpha$), the resulting temperature at the outer
radius would be higher than 25 K if we use the same value for $T_{0}$.
In that case, we adjusted $T_{0}$ such that \mbox{$T(R_{\mathrm{out}})
= 25$ K}. This outer boundary temperature is in the regime of
excitation temperatures of the lower rotational transitions.  For the
negative values of $\alpha$, the temperature $T_{0}$ was assumed to be
25 K. A number of models with $\alpha = 0$ has also been computed, see
the lower left panel of Fig.~\ref{fig:mosaic2}. Adopted temperatures
are 10, 25, 50 and 100 K.  The models with a constant temperature or
an outwards increasing temperature are unrealistic, but we included
them in our parameter study, to see if it is possible at all to change
the line ratios significantly by changing the run of the temperature.

\begin{table}
\caption{Parameters for the temperature profiles with a change in slope. 
See text for details.}
\begin{center}
\begin{tabular}{l l l l}
\hline
\hline
profile           & $\alpha_{\mathrm{in}}$  & $T_0$ (K) & $T_{\mathrm{ex}}$ (K)\\  
\hline
1                  & 1.0         & 2500      & 33.1 \\
2                  & 1.5         & 2500      & 33.1\\
3                  & 1.0         & 2500      & 16.6\\
4                  & 1.0         & 2500      & 55.2\\
5                  & 1.0         & 2200      & 33.1\\
\hline
\hline
\end{tabular}
\end{center}
\label{tab:kink}
\end{table}

To add to the realism of the models, we composed a number of temperature
profiles consisting of two power laws with different values for
$\alpha$. These profile are inspired by heat balance calculations
of \citet{JCT_94_redgiantwinds}, and are defined as follows:
\begin{displaymath}
  T(r) = \left\{ \begin{array}{ll} 
      T_0 (r/R_{\ast})^{-\alpha_{\mathrm{in}}} &\mathrm{for} \, T > T_{\mathrm{ex}}\\
      T_1 (r/R_{\ast})^{-\alpha_{\mathrm{out}}} &\mathrm{for} \, T < T_{\mathrm{ex}}
    \end{array} \right.
\end{displaymath}
Five different profiles with a change in slope were constructed, where
the excitation temperatures of the CO(4$-$3) ($T_{\mathrm{ex}}=55.2$
K), (3$-$2) (33.1 K) and (2$-$1) (16.6 K) were used to define the
position of the change in the slope.  In all cases
$\alpha_{\mathrm{out}}$ was chosen to be 0.7. For the other
parameters, the reader is referred to Table~\ref{tab:kink}. The
resulting line ratios are plotted in the upper right panel of
Fig.~\ref{fig:mosaic2}. All models with a power law with a slope
change cluster remarkably close to the CO(4$-$3)/CO(2$-$1) and
CO(3$-$2)/CO(2$-$1) ratios observed in \object{CRL 2199} and
\object{WX Psc}.  The only outlier is profile 2 (see
Table~\ref{tab:kink}). Although from the various panels in
Fig.~\ref{fig:mosaic2} it seems to be possible to explain the observed
ratios of \object{WX Psc} and \object{CRL 2199}, 
it is not possible to explain the line ratios of
other stars of our sample, not even for extreme temperature profiles.

\subsection{A representative case: WX Psc}
\label{sec:wxpsc}

In order to investigate the possibilities to explain the integrated
intensities of the CO rotational transitions, we will focus on
\object{WX Psc}. All transitions are observed and detected. The
signal-to-noise ratio is reasonable for all transition, except for the
CO(6$-$5) line.  The previous
section has shown that the line ratios of the lower rotational
transitions can be explained using power law temperature profiles. In
this section we expand our investigations to the higher rotational
transitions.  The observed values for the integrated intensities $I$
of the CO(6$-$5) and (7$-$6) transition are much lower than the
expected values based on the standard model described in
Sect.~\ref{sec:analysis}.

\begin{table}
\caption{Physical parameters of \object{WX Psc}. The terminal velocity 
  $v_{\infty}$ and system velocity of the object $v_{\mathrm{LSR}}$ are 
  derived from our observations (Table~\ref{tab:transitions}). The other 
  parameters are extracted from the literature, the references are: 
$^{1}$\citet{VVV_90_phaselags}, $^{2}$\citet{HBB_01_WXPsc}, 
$^{3}$\citet{JCT_94_redgiantwinds}, $^{4}$\citet{ZE_00_WHya}, 
$^{5}${\sc simbad}, $^{6}$\citet{LL_96_SED}, 
$^{7}$\citet{LW_00_coolstars}.}
\begin{center}
\begin{tabular}{l l c}
\hline
\hline
parameter          & value                          & ref.\\
\hline
distance           & 0.74 kpc                       & 1 \\
$v_{\infty}$       & 20 km s$^{-1}$                 &\\
$v_{\mathrm{LSR}}$ & +9 km s$^{-1}$                 &\\
$R_{\mathrm{in}}$  & 6.6 R$_{\ast}$                 & 2 \\
$T_{\mathrm{eff}}$ & 2250 K                         & 2 \\
$\alpha$           & 0.5                            & 3,4 \\
sp.~type           & M9-10                          & 5 \\
$L_{\ast}$         & $(1.22-1.31) \cdot 10^4$ L$_{\odot}$  & 6 \\
$M_{\ast}$         & $>5$ M$_{\odot}$               & 7 \\
\hline
\hline
\end{tabular}
\end{center}
\label{tab:wxpsc}
\end{table}

\object{WX Psc} is a well studied AGB star with an intermediate
mass-loss rate. From recent studies, notably the work of
\citet{HBB_01_WXPsc}, we have retrieved the physical characteristics
of the star and the circumstellar environment (see
Table~\ref{tab:wxpsc}). These values were used as input parameters for
our model calculations. For required parameters which are not
accurately known, we maintained the values of our standard model
(Table~\ref{tab:standard}).  We used a stellar radius of $5.4 \cdot
10^{13}$ cm, implying a luminosity of $1.3 \cdot 10^4$
L$_{\odot}$ for $T_{\mathrm{eff}}=2250$ K.

\begin{table*}
\caption{Mass-loss rates for \object{WX Psc}. The values are derived for each
observed transition independently. While all other 
parameters of \object{WX Psc} were kept constant, the mass-loss rate was
determined by fitting the integrated intensity. The third column contains the excitation 
temperature of the corresponding rotational transition. 
The fourth and fifth column represent the outflow distances, i.e.~the distance 
traveled since the gas left the stellar surface, and the
last column shows the corresponding travel times. 
In addition to the CO mass-loss determinations, 
the mass-loss rate derived from
the $L-[12\,\mu\mathrm{m}]$ colour, assuming a dust-to-gas ratio of 0.01, is given \citep{KDW_02_composition}.}
\begin{center}
\begin{tabular}{l c l r c r}
\hline
\hline
tracer & $\dot{M}_{\mathrm{gas}}$ & $T_{\mathrm{ex}}$ (K) & $R/R_{\ast}$ & $R$ (cm) & travel time \\
                            & (M$_{\odot}$ yr$^{-1}$)       &  &&  & (yr)\\
\hline
$L - [12\, \mu \mathrm{m}]$ & $2.0 \cdot 10^{-5}$           &      &      &                     & $<600$\\
CO 7--6                     & $3.0 (\pm 0.3) \cdot 10^{-7}$ & 155  & 700  & $3.8 \cdot 10^{16}$ & 600\\
CO 6--5                     & $1.4 (\pm 0.1) \cdot 10^{-7}$ & 116  & 900  & $4.9 \cdot 10^{16}$ & 780\\
CO 4--3                     & $1.3 (\pm 0.1) \cdot 10^{-6}$ & 55.1 & 1400 & $7.6 \cdot 10^{16}$ & 1200\\
CO 3--2                     & $6.3 (\pm 0.2) \cdot 10^{-6}$ & 33.1 & 1100 & $5.9 \cdot 10^{16}$ & 940\\
CO 2--1                     & $8.0 (\pm 0.9) \cdot 10^{-6}$ & 16.6 & 1900 & $1.0 \cdot 10^{17}$ & 1600\\ 
\hline
\hline
\end{tabular}
\end{center}
\label{tab:wxmdot}
\end{table*}

\begin{figure}
  \centerline{\includegraphics[width=8cm]{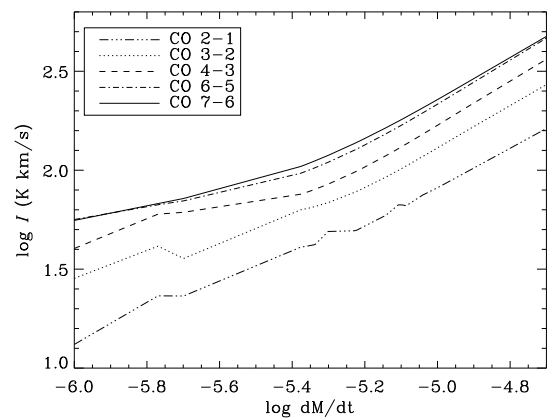}}
  \caption{Using the known parameters of \object{WX Psc}
    (Table~\ref{tab:wxpsc}) predicted integrated intensities are given
    for a large range of mass-loss rates. Integrated intensities
    are plotted in a logarithmic scale on the vertical axis, and 
    mass-loss rates ($M_{\odot}$ yr$^{-1}$)  on the horizontal
    axis, also in logarithmic scale. The integrated intensities have
    been calculated for all lines observable with the JCMT and these
    calculated models are indicated with symbols (see legend). The
    models have been connected with a line. Using the observed
    integrated intensity for a certain line, the mass-loss rate of WX
    Psc can be estimated from this plot (see Table~\ref{tab:wxmdot}).}
  \label{fig:ewdmdt}
\end{figure}

Determination of the mass-loss rate from the integrated intensities of
the rotational transitions gives an idea of the mass-loss
history of the AGB star. Fig.~\ref{fig:ewdmdt} shows how the
integrated line intensities depend on it. In
Table~\ref{tab:wxmdot} the mass-loss rates derived from 
each observed transition are listed, while all other parameters
were kept fixed. In addition, the gas mass-loss rate, derived from the
$L - [12\, \mu \mathrm{m}]$ colour \citep{KDW_02_composition} is given,
where a dust-to-gas ratio of 1\% is assumed. The dust spectral energy
distribution covers a temperature range of $\sim200-800$ K, which
corresponds to a region even more inwards than the CO line emission.

We conclude that constant mass-loss rate models cannot explain all of
the observed line intensities. Rather, it seems that the mass-loss
rate varies with the $J$-level under consideration.  Specifically, the
mass-loss rate corresponding to the CO(2$-$1) emission is almost
comparable in strength to the mass-loss rate derived from the dust
emission (Table~\ref{tab:wxmdot}).  For the higher rotational
transitions, the derived mass-loss rates go down with increasing line
frequency, although it perhaps increases slightly again for the
CO(7$-$6) transition.  The mass-loss rates determined from the high
rotational transitions disagree with the mass-loss rate derived from
the infrared dust emission. A difference of at least an order of
magnitude occurs although the regions that are traced by the high
rotational transitions and the dust emission are closest in
temperature, and are therefore spatially close together.  In general,
a decreasing mass-loss rate with increasing rotational energy level is
observed, which is inconsistent with predictions based on the
superwind model
\citep[e.g.][]{G_94_OH32OH44,JST_96_OH26,DKF_97_superwind}.  The
results derived here point towards a mass-loss rate decreasing with
time, rather than a stratification consistent with the onset of a
superwind phase. In the next sections we will try to explain this
discrepancy.

\subsection{Possible explanations for the inconsistency}
\label{sec:expl}

\subsubsection{Mass-loss variations?}

\begin{figure}
  \centerline{\includegraphics[width=8cm]{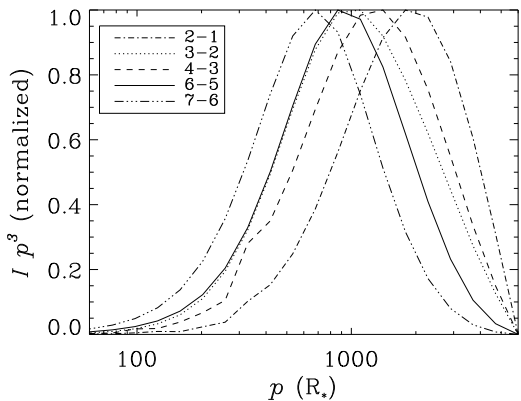}}
  \caption{Normalized intensity ($I(p) p^3$) as a function of 
    impact parameter $p$.  The curves for each rotational transition
    are calculated using the mass-loss rate corresponding to that
    transition (see Table~\ref{tab:wxmdot}). }
  \label{fig:reg}
\end{figure}

In principle, it should be possible to construct a \emph{combination
  of a density and temperature profile}, such that the observed line
  intensity ratios can be explained. This is not possible for a
  constant mass-loss rate, as becomes apparent from
  Fig.~\ref{fig:mosaic2}, so apparently there must also have been
  variations in $\dot{M}$.  Using the \object{WX Psc} model
  calculations, we can derive an impression of the mass-loss history
  using the mass-loss rates listed in Table~\ref{tab:wxmdot}.  For all
  these values we have calculated the region where the respective line
  originates (Fig.~\ref{fig:reg}). This is not done in terms of radial
  distance to the central star, but as a function of impact parameter,
  in which case contributions in line-of-sights due to interactions at
  various radial distances have been integrated. Therefore, the values
  on the horizontal axis can not directly be translated to a radial
  distance towards the central star, but present a lower limit to this
  distance.  In addition, one has to bear in mind that the regions
  from which the various lines originate are not distinct, but largely
  overlap. Some overlap in Fig.~\ref{fig:reg} is due to projection
  effects along the line-of-sight, but a significant fraction is due
  to real physical overlap of the line-formation regions.  Although
  all regions are plotted in one figure, they do not arise from the
  same model but are the calculated for the corresponding mass-loss
  rate for each line (see Table~\ref{tab:wxmdot}).  Therefore, it is
  possible that the CO (4$-$3) seems to originate from a region that
  is more distant from the central star than the region where the CO
  (3$-$2) line originates, although their excitation temperatures
  would suggest otherwise in an outwards decreasing temperature
  profile.  Concluding, the mass-loss rates that we have determined
  are only average values for these line formation regions.
  Nevertheless, estimates of the distances from the line forming
  regions towards the central star can be derived for the mass-loss
  rates traced by the observed transitions.  Using a stellar radius of
  $5.4 \cdot 10^{13}$ cm and an expansion velocity of 20 km s$^{-1}$
  the time elapsed since the ejection of the gas from the stellar
  surface, traced by the various transitions can be calculated.  The
  results are listed in Table~\ref{tab:wxmdot}. The cycle can be
  completed by adding the dust mass-loss, mostly originating from the
  region inwards of the CO(7$-$6) transition, and thus ejected less
  than 600 years ago.  Note that the dust mass-loss rate is
  transferred into a gas mass-loss rate by assuming a dust-to-gas mass
  ratio of 0.01. Actual deviations to this ratio imply a different gas
  mass-loss rate traced by the $L-[12$ $\mu$m$]$ colour.  From
  Table~\ref{tab:wxmdot} we can determine that the interval between the
  two maximum mass-loss rates, traced by the CO(2$-$1) transition and
  the $L-[12$ $\mu$m$]$ colour of the dust emission, is of the order
  of $\sim$1000 years.

Mass-loss variations on such time scales have in fact been observed in
other evolved stars.  Circumstellar series of arc-like structures have
been interpreted as due to mass-loss modulations, notably for C-rich
post-AGB stars, where the separation is a measure for the time scale
of these variations.  \citet{KSH_98_bipolar} derive that the
separation between arcs observed around {IRAS 17150$-$3224}
corresponds to a time scale of \mbox{240 yr ($D$/kpc)
($v_{\mathrm{exp}}$/10 km s$^{-1}$)}. For reasonable numbers for the
distance and outflow velocity one can determine that these arcs may be
due to mass-loss variations on time scales of 200--1000 yr. A similar
time scale (200--800 yr) is derived by \citet{MH_99_IRC10216} for
{IRC+10216}. The circumstellar arcs around {CRL 2688} (Egg Nebula) are
believed to be ejected at 75--200 yr intervals, assuming a distance of
1 kpc and an outflow velocity of 20 km s$^{-1}$
\citep{STW_98_eggnebula}. IRAS LRS spectroscopy has shown that hot
dust ($T>500$ K) is absent around a number of AGB stars. This is
interpreted as a drop in mass-loss rate which occurred $\sim$100 years
ago, consistent with the spacing between the arcs observed around
post-AGB stars \citep{MIZ_01_100yr}.  Hydrodynamic calculations
considering the gas and dust as partially or completely decoupled
outflow components resulted in mass-loss variations of an order of
magnitude at intervals of 200--350 year for partially and 400 year for
completely decoupled fluids \citep{SID_01_quasiperiodic}.  Moreover,
\citet{FMS_03_multiple} report on the discovery of multiple shells
seen in CO (1$-$0) emission around \object{IRC+10216}. These shells
are found to have intershell time scales of 1300--2900 year. The
circumstellar arcs and molecular shells observed around post-AGB stars
and the density enhancements emerging from hydrodynamic calculations
have similar time scales to what we derive here for mass-loss
variations in the outflow of \object{WX Psc}, indicating that the same
phenomenon may perhaps play a role here.

Variations in the mass-loss rate of AGB stars have already been
studied for a long time.  It is generally accepted that the AGB phase
is terminated by the superwind; a phase in which the mass-loss rate
rapidly increases \citep{R_81_superwind,BH_83_maserstrength}. However,
the mass-loss rates inferred from the CO line intensities for
\object{WX~Psc} decrease with time and are thus opposite to the
classical superwind model predictions.  The thermal pulses associated
with He-shell ignition are also thought to cause mass-loss variations
\citep{VW_93_massloss}. As for the superwind, the behaviour and time
scale of these variations do not comply with our model predictions.

\subsubsection{A gradient in the turbulent velocity?}

\begin{figure}
  \centerline{\includegraphics[width=8cm]{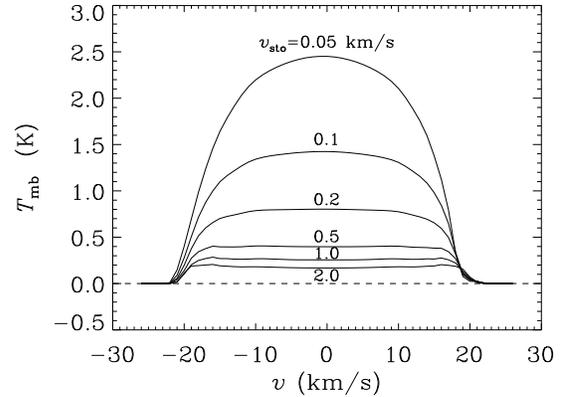}}
  \caption{The influence on the line profile of CO(3$-$2) due to variations 
    of the stochastic velocity. The input parameters of our standard
    model are used (Table~\ref{tab:standard}), only the turbulent
    velocity -- which in our model is independent of $r$ -- is varied,
    in the range from 0.05 -- 2.0 km s$^{-1}$.}
  \label{fig:vsto}
\end{figure}

Besides a complex density-temperature profile due to periodic
mass-loss variations, there may be another way to explain the line
intensities of the CO rotational transitions observed in \object{WX
  Psc}; a \emph{gradient in stochastic velocity $v_{\mathrm{sto}}$}.
The stochastic velocity is an important parameter in the
line formation process (see Sect.~\ref{sec:conditions}).
Fig.~\ref{fig:vsto} illustrates this for the profiles of one of the
rotational transitions for various stochastic velocities.

\begin{table}
\caption{Stochastic velocities for \object{WX Psc}. The values are 
derived for each
observed transition independently, while all other 
parameters were kept constant. The stochastic velocities were determined by 
fitting the integrated intensities. }
\begin{center}
\begin{tabular}{l l}
\hline
\hline
tracer          & $v_{\mathrm{sto}}$\\
                   & (km s$^{-1}$)    \\
\hline
CO(7$-$6)            & 3.2 $\pm$ 0.4             \\  
CO(6$-$5)            & 8 $\pm$ 1             \\  
CO(4$-$3)            & 1.0 $\pm$ 0.1              \\ 
CO(3$-$2)            & 0.24 $\pm$ 0.04             \\ 
CO(2$-$1)            & 0.16 $\pm$ 0.05              \\
\hline
\hline
\end{tabular}
\end{center}
\label{tab:vsto}
\end{table}

Analogous to the determination of the mass-loss history, it is
possible to estimate the variations in the stochastic velocities
traced by the integrated intensities of the sequence of rotational
transition observed for \object{WX~Psc}. For this purpose, the
mass-loss rate was assumed to be constant at a rate of 10$^{-6}$
$M_{\odot}$ yr$^{-1}$ throughout the circumstellar outflow. The
temperature profile and the other parameters were kept the same as the
ones used in the mass-loss history analysis. The results are listed in
Table~\ref{tab:vsto}. Again, the line formation regions and thus the
values derived here are not independent and should be seen as averages
over the formation regions.  The observed line intensities may be
explained by a gradient in the stochastic velocity if it is lowest in
the outer parts of the outflow, traced by the low rotational
transitions, and has its maximum in the gas traced by the CO(6$-$5)
transition.

One has to bear in mind that we derived the stochastic velocities for
one particular mass-loss rate, namely $10^{-6} \, M_{\odot}$. As
pointed out before, the mass-loss rate has a considerable effect on
the line strengths as well, however, the negative gradient will be
maintained for other choices of $\dot{M}$.  To explain the
observations $v_{\mathrm{sto}}$ has to increase to an unrealistically
high maximum of 8 km s$^{-1}$ in the region of CO(6$-$5) formation and
then decrease again to 0.16 km s$^{-1}$ at the CO(2$-$1) formation
zone.

The stochastic velocity can be considered as a composition of thermal
and turbulent components, according to Eq.~(\ref{eq:velocities}).  The
thermal molecular velocities for CO are given by ${v_{\mathrm{therm}}}
= \sqrt{2 k T / m_{\mathrm{CO}}}$, where $T$ is the temperature at the
line formation region, which is usually of the order of the excitation
temperature. For the lines observed we can determine that
$v_{\mathrm{therm}}$ ranges between 0.01 and 0.03 km s$^{-1}$. It is
obvious that only a minor fraction of the required total stochastic
velocity ${v_{\mathrm{sto}}}$ can be explained by thermal motion, and
also that the observed gradient is not sufficiently reproduced by the
thermal component.

The nature of the remaining turbulent velocity $v_{\mathrm{turb}}$ is
unknown, but could in part be induced by stellar pulsations.  These
pulsations cause stochastic velocities of 2--5 km s$^{-1}$ in the
inner parts of the circumstellar shell required to start the dust
formation process.  However, these stochastic velocities will damp
quickly and are practically absent beyond \mbox{100 $R_{\ast}$}
\citep{S_01_modulation}. Hence, if variations in the stochastic
velocity are important for the CO line intensities, the origin of such
variations is presently unclear.

\subsubsection{Other factors}

Other factors that could be important include the outflow velocity
profile, and the geometry. Our adopted outflow velocity profile is
very simple (see Eq.~(\ref{eq:vel})), but hydrodynamical calculations
show that it may be more complex and also time-dependent
\citep{SID_01_quasiperiodic}. The effect of such complex outflow
velocity profiles on the CO line profiles has not yet been
studied. Perhaps they could serve as a source of turbulence.

Non-spherical winds, e.g.~density enhancements in the equatorial
region, could also play a role in the observed line strengths. The
observed line profiles would reflect such an axi-symmetric geometry,
if it exists. Close examination of the observed lines
(Figs.~\ref{fig:wxpsc}--\ref{fig:oh104}) shows that their profiles are
similar for all transitions (per source), and one can therefore
conclude that the regions where the lines originate have almost the
same velocity structure.  Apparently there is no change in geometry
for the regions traced by the various rotational transitions, e.g.~a
slowly outflowing disk traced by the lower transitions and a fast
polar outflow traced by the higher transitions. Thus this possibility
most likely can be ruled out as an explanation for the
discrepancy between the observations and the model results.  Only in
case of \object{VY CMa} (Fig.~\ref{fig:vycma}) the profiles show
significant differences between the lower and higher rotational
transitions.

\section{Concluding remarks}
\label{sec:disc}

\subsection{CO rotational transitions as mass-loss indicators}
\label{sec:co}

In this work, we presented submillimeter observations of various
carbon monoxide rotational transitions (CO(7$-$6), (6$-$5), (4$-$3),
(3$-$2), (2$-$1)) observed in AGB stars and red supergiants in various
evolutionary states.  We have attempted to determine the mass-loss
history of the programme stars by modelling of the observed
transitions.  For the first time the CO(7$-$6) and (6$-$5) transitions
were used, in addition to lower transitions. In this way the gap
between the regions in the outflow traced by the gas and that traced
by the dust emission was largely closed.  Many studies in the past
have focussed on only one or two transitions to determine the gas
mass-loss rate
\citep[e.g.][]{KM_85_massloss,LFO_93_CO,JCT_94_redgiantwinds,G_94_OH32OH44,JST_96_OH26,DKF_97_superwind}.
The extension of the data towards higher rotational transitions
clearly demonstrates that determination of a unique gas mass-loss rate
from a single CO rotational transition is highly unreliable.  We found
that the observed line strengths indicate that the outflow has a more
complex physical structure than was previously assumed. Not a
superwind, but periodic mass-loss variations comparable to the
arc-like structures and rings observed around post-AGB stars, may
possibly account for the observed line strengths. Part of the
discrepancy could be due to a gradient in the stochastic velocity as
well.

Independently, another research group has reached the same
  conclusion during the last year.  Initially, \citet{OGK_02_massloss}
  modelled the mass-loss rates of a large sample of irregular and
  semi-regular M-type variables by fitting 2, 3 or in one case 4 CO
  rotational transitions by assuming a constant mass-loss rate over
  the last 1000 years.  They derive rates for their sample stars and
  do not report on problems similar to ours, but their figures 2 and
  11 show that the line strength of the higher transitions is
  overestimated when this model is used.  In the same volume of A\&A,
  \citet{SRO_02_mdothistory} describe a model that is able to use
  periodic mass-loss variations to calculate the rotational
  transitions of CO in C-rich stars. The development of this model is
  driven by the discovery of mass-loss modulations. However, after
  thorough analysis, they conclude that mass-loss modulations are not
  important nor necessary to explain the CO rotational line profiles.
  The most recent results of \citet{GOK_03_dynamics} indicate
  otherwise, however. When trying to derive the mass loss rate of more
  evolved Miras (i.e.~with higher mass-loss rates than the
  semi-regulars), \citet{GOK_03_dynamics} find that a model assuming a
  constant mass-loss rate underestimates the strength of the low
  transitions. This is in principle the same as our result that the
  high transitions are overestimated.  

\subsection{Future work}
\label{sec:future}

The work presented here has revealed a much more complex picture of
AGB stellar ejecta than previously assumed. Additional research is required,
which we plan to do in the near future. Of particular importance are
the following issues:

\begin{itemize}
  
\item First, more observational data should be obtained, in particular
  of high rotational transitions. Our study is the first to include
  the CO(6$-$5) and (7$-$6) transitions in the mass-loss rate
  determinations of three evolved stars. In addition, for one object
  (\object{NML Cyg}) observations up to CO(6$-$5) were secured. This is not
  enough to draw firm conclusions on the degree of complexity of the
  physical structure in the outflows of AGB stars, therefore this
  sample should be enlarged.  It is important to pay attention to the
  completeness: if all transitions are observed, variations in the
  important physical parameters can be much better constrained.  In
  that respect it is also worthwhile to extend the data with
  observations of $^{13}$CO for the lower transitions, which provide
  additional independent constraints on the physical conditions.
  
\item Second, a more realistic representation of the physical
  conditions in the outflow of AGB stars should be used. This includes
  adding a gradient in turbulence and periodic mass-loss variations as
  we have argued in this study. In addition, the velocity law could
  also be improved, e.g.~following the results of
  \citet{SID_01_quasiperiodic}. Although these adjustments will lead
  to an increase in the number of free parameters, it is likely that
  we will be able to use the \emph{line profiles} to constrain the
  model parameters. This will certainly help in disentangling the
  physical structure of the outflow.
  
\end{itemize}

\begin{acknowledgements}
  FK is grateful for the hospitality at the Stockholm Observatory.
  The support provided by the staff of the JCMT is greatly
  appreciated.  The James Clerk Maxwell Telescope is operated on
  behalf of the Particle Physics and Astronomy Research Council of the
  United Kingdom, the Netherlands Organisation for Scientific Research
  and the National Research Council of Canada. This research has made
  use of the SIMBAD database, operated at CDS, Strasbourg, France. FK,
  AdK and LBFMW acknowledge financial support from NWO 'Pionier' grant
  616-78-333. Support for this work was provided by NASA through the
  SIRTF Fellowship Program, under award 011 808-001.
  
\end{acknowledgements}

\bibliographystyle{aa} 
\bibliography{ciska}

\begin{thebibliography}{61}
\expandafter\ifx\csname natexlab\endcsname\relax\def\natexlab#1{#1}\fi

\bibitem[{Baud \& Habing(1983)}]{BH_83_maserstrength}
Baud, B. \& Habing, H.~J. 1983, \aap, 127, 73

\bibitem[{Bedijn(1987)}]{B_87_dustshells}
Bedijn, P.~J. 1987, \aap, 186, 136

\bibitem[{Bergman {et~al.}(2000)Bergman, Kerschbaum, \&
  Olofsson}]{BKO_00_rvboo}
Bergman, P., Kerschbaum, F., \& Olofsson, H. 2000, \aap, 353, 257

\bibitem[{Cami {et~al.}(1998)Cami, {de Jong}, Justtanont, Yamamura, \&
  Waters}]{CJJ_98_ohir}
Cami, J., {de Jong}, T., Justtanont, K., Yamamura, I., \& Waters, L. B. F.~M.
  1998, \apss, 255, 339

\bibitem[{Chen \& Neufeld(1995)}]{CN_95_water}
Chen, W. \& Neufeld, D.~A. 1995, \apj, 453, L99

\bibitem[{Danchi {et~al.}(2001)Danchi, Green, Hale, McElroy, Monnier, Tuthill,
  \& Townes}]{DGH_01_NMLCyg}
Danchi, W.~C., Green, W.~H., Hale, D. D.~S., {et~al.} 2001, \apj, 555, 405

\bibitem[{Delfosse {et~al.}(1997)Delfosse, Kahane, \&
  Forveille}]{DKF_97_superwind}
Delfosse, X., Kahane, C., \& Forveille, T. 1997, \aap, 320, 249

\bibitem[{Dijkstra {et~al.}(2003)Dijkstra, Waters, Kemper, Zijlstra, Matsuura,
  {de Koter}, Hony, \& Dominik}]{DWK_02_IRAS16342}
Dijkstra, C., Waters, L. B. F.~M., Kemper, F., {et~al.} 2003, \aap, 399, 1037

\bibitem[{Flower \& Launay(1985)}]{FL_85_ratecoefficients}
Flower, D.~R. \& Launay, J.~M. 1985, \mnras, 214, 271

\bibitem[{Fong {et~al.}(2002)Fong, Justtanont, Meixner, \&
  Campbell}]{FJM_02_oh26}
Fong, D., Justtanont, K., Meixner, M., \& Campbell, M.~T. 2002, \aap, 396, 581

\bibitem[{Fong {et~al.}(2003)Fong, Meixner, \& Shah}]{FMS_03_multiple}
Fong, D., Meixner, M., \& Shah, R.~Y. 2003, \apjl, 582, L39

\bibitem[{Gehrz(1989)}]{G_89_stardust}
Gehrz, R. 1989, in IAU Symp.~135: Interstellar Dust, ed. L.~J. Allamandola \&
  A.~G. G.~M. Tielens, 445--453

\bibitem[{Goldreich \& Scoville(1976)}]{GS_76_OHIR}
Goldreich, P. \& Scoville, N. 1976, \apj, 205, 144

\bibitem[{{Gonz\'alez Delgado} {et~al.}(2003){Gonz\'alez Delgado}, Olofsson,
  Kerschbaum, Sch\"oier, \& Lindqvist}]{GOK_03_dynamics}
{Gonz\'alez Delgado}, D., Olofsson, H., Kerschbaum, F., Sch\"oier, F.~L., \&
  Lindqvist, M. 2003, \aap, in press

\bibitem[{Groenewegen(1994{\natexlab{a}})}]{G_94_revisedmodel}
Groenewegen, M. A.~T. 1994{\natexlab{a}}, \aap, 290, 531

\bibitem[{Groenewegen(1994{\natexlab{b}})}]{G_94_OH32OH44}
---. 1994{\natexlab{b}}, \aap, 290, 544

\bibitem[{Habing(1996)}]{H_96_review}
Habing, H. 1996, \aapr, 7, 97

\bibitem[{Herman {et~al.}(1986)Herman, Burger, \& Penninx}]{HBP_86_luminosity}
Herman, J., Burger, J.~H., \& Penninx, W.~H. 1986, \aap, 167, 247

\bibitem[{Heske {et~al.}(1990)Heske, Forveille, Omont, {van der Veen}, \&
  Habing}]{HFO_90_deficiency}
Heske, A., Forveille, T., Omont, A., {van der Veen}, W. E. C.~J., \& Habing,
  H.~J. 1990, \aap, 239, 173

\bibitem[{Hofmann {et~al.}(2001)Hofmann, Balega, Bl\"ocker, \&
  Weigelt}]{HBB_01_WXPsc}
Hofmann, K.-H., Balega, Y., Bl\"ocker, T., \& Weigelt, G. 2001, \aap, 379, 529

\bibitem[{Hyland {et~al.}(1972)Hyland, Becklin, Frogel, \&
  Neugebauer}]{HBF_72_1612MHz}
Hyland, A.~R., Becklin, E.~E., Frogel, J.~A., \& Neugebauer, G. 1972, \aap, 16,
  204

\bibitem[{{Iben Jr.} \& Renzini(1983)}]{IR_83_AGBevolution}
{Iben Jr.}, I. \& Renzini, A. 1983, \araa, 21, 271

\bibitem[{Jackson {et~al.}(2000)Jackson, {de Lange}, Laauwen, Gao, Iosad, \&
  Klapwijk}]{JDL_00_E}
Jackson, B.~D., {de Lange}, G., Laauwen, W.~M., {et~al.} 2000, in Proc.~of the
  11th Int.~Symp.~on Space THz Technology, 238

\bibitem[{Jones {et~al.}(1993)Jones, Humphreys, Gehrz, Lawrence, Zickgraf,
  Moseley, Casey, Glaccum, Koch, Pina, Jones, Venn, Stahl, \&
  Starrfield}]{JHG_93_irc+10420}
Jones, T.~J., Humphreys, R.~M., Gehrz, R.~D., {et~al.} 1993, \apj, 411, 323

\bibitem[{Justtanont {et~al.}(1994)Justtanont, Skinner, \&
  Tielens}]{JCT_94_redgiantwinds}
Justtanont, K., Skinner, C.~J., \& Tielens, A. G. G.~M. 1994, \apj, 435, 852

\bibitem[{Justtanont {et~al.}(1996)Justtanont, Skinner, Tielens, Meixner, \&
  Baas}]{JST_96_OH26}
Justtanont, K., Skinner, C.~J., Tielens, A. G. G.~M., Meixner, M., \& Baas, F.
  1996, \apj, 456, 337

\bibitem[{Kahane \& Jura(1996)}]{KJ_96_XHer}
Kahane, C. \& Jura, M. 1996, \aap, 310, 952

\bibitem[{Kemper {et~al.}(2002)Kemper, {de Koter}, Waters, Bouwman, \&
  Tielens}]{KDW_02_composition}
Kemper, F., {de Koter}, A., Waters, L. B. F.~M., Bouwman, J., \& Tielens, A. G.
  G.~M. 2002, \aap, 384, 585

\bibitem[{Kemper {et~al.}(2001)Kemper, Waters, {de Koter}, \&
  Tielens}]{KWD_01_xsilvsmdot}
Kemper, F., Waters, L. B. F.~M., {de Koter}, A., \& Tielens, A. G. G.~M. 2001,
  \aap, 369, 132

\bibitem[{Kerschbaum \& Olofsson(1999)}]{KO_99_COcatalogue}
Kerschbaum, F. \& Olofsson, H. 1999, \aaps, 138, 299

\bibitem[{Kerschbaum {et~al.}(2003)Kerschbaum, Olofsson, Posch, {Gonz\'alez
  Delgado}, Bergman, Mutschke, J\"ager, Dorschner, \&
  Sch\"oier}]{KOP_03_massloss}
Kerschbaum, F., Olofsson, H., Posch, T., {et~al.} 2003, \rma, in press

\bibitem[{Knapp \& Morris(1985)}]{KM_85_massloss}
Knapp, G.~R. \& Morris, M. 1985, \apj, 292, 640

\bibitem[{Knapp {et~al.}(1998)Knapp, Young, Lee, \& Jorissen}]{KYL_98_COsurvey}
Knapp, G.~R., Young, K., Lee, E., \& Jorissen, A. 1998, \apjs, 117, 209

\bibitem[{Kwok {et~al.}(1998)Kwok, Su, \& Hrivnak}]{KSH_98_bipolar}
Kwok, S., Su, K. Y.~L., \& Hrivnak, B.~J. 1998, \apjl, 501, L117

\bibitem[{Lan{\c c}on \& Wood(2000)}]{LW_00_coolstars}
Lan{\c c}on, A. \& Wood, P.~R. 2000, \aaps, 146, 217

\bibitem[{{Le Sidaner} \& {Le Bertre}(1996)}]{LL_96_SED}
{Le Sidaner}, P. \& {Le Bertre}, T. 1996, \aap, 314, 896

\bibitem[{Loup {et~al.}(1993)Loup, Forveille, Omont, \& Paul}]{LFO_93_CO}
Loup, C., Forveille, T., Omont, A., \& Paul, J.~F. 1993, \aaps, 99, 291

\bibitem[{Marengo {et~al.}(2001)Marengo, Ivezi\'c, \& Knapp}]{MIZ_01_100yr}
Marengo, M., Ivezi\'c, {\v Z}., \& Knapp, G.~R. 2001, \mnras, 324, 1117

\bibitem[{Mauron \& Huggins(1999)}]{MH_99_IRC10216}
Mauron, N. \& Huggins, P.~J. 1999, \aap, 349, 203

\bibitem[{Mihalas {et~al.}(1975)Mihalas, Kunasz, \& Hummer}]{MKH_75_twolevel}
Mihalas, D., Kunasz, P.~B., \& Hummer, D.~G. 1975, \apj, 202, 465

\bibitem[{Molster {et~al.}(2002)Molster, Waters, Tielens, \&
  Barlow}]{MWT_02_xsilI}
Molster, F.~J., Waters, L. B. F.~M., Tielens, A. G. G.~M., \& Barlow, M.~J.
  2002, \aap, 382, 184

\bibitem[{Morris(1980)}]{M_80_CO}
Morris, M. 1980, \apj, 236, 823

\bibitem[{Olofsson {et~al.}(2002)Olofsson, {Gonz\'alez Delgado}, Kerschbaum, \&
  Sch\"oier}]{OGK_02_massloss}
Olofsson, H., {Gonz\'alez Delgado}, D., Kerschbaum, F., \& Sch\"oier, F.~L.
  2002, \aap, 391, 1053

\bibitem[{Perryman \& ESA(1997)}]{ESA_97_Hipparcos}
Perryman, M.~A.~C. \& ESA. 1997, {The HIPPARCOS and TYCHO catalogues.
  Astrometric and photometric star catalogues derived from the ESA HIPPARCOS
  Space Astrometry Mission}, ESA-SP 1200 (Noordwijk, The Netherlands: ESA
  Publications Division)

\bibitem[{Renzini(1981)}]{R_81_superwind}
Renzini, A. 1981, in Physical processes in red giants, ed. I.~{Iben Jr.} \&
  A.~Renzini (Dordrecht, Holland: D. Reidel Publishing Company), 431--446

\bibitem[{Sahai {et~al.}(1998)Sahai, Trauger, Watson, Stapelfeldt, Hester,
  Burrows, Ballister, Clarke, Crisp, Evans, {Gallagher III}, Griffiths,
  Hoessel, Holtzman, Mould, Scowen, \& Westphal}]{STW_98_eggnebula}
Sahai, R., Trauger, J.~T., Watson, A.~M., {et~al.} 1998, \apj, 493, 301

\bibitem[{Sch\"oier {et~al.}(2002)Sch\"oier, Ryde, \&
  Olofsson}]{SRO_02_mdothistory}
Sch\"oier, F.~L., Ryde, N., \& Olofsson, H. 2002, \aap, 391, 577

\bibitem[{Sch\"onberg(1988)}]{S_88_molspec}
Sch\"onberg, K. 1988, \aap, 195, 198

\bibitem[{Sch\"onberg \& Hempe(1986)}]{SH_86_multilevel}
Sch\"onberg, K. \& Hempe, K. 1986, \aap, 163, 151

\bibitem[{Simis(2001)}]{S_01_modulation}
Simis, Y. J.~W. 2001, PhD thesis, Leiden University, The Netherlands

\bibitem[{Simis {et~al.}(2001)Simis, Icke, \& Dominik}]{SID_01_quasiperiodic}
Simis, Y. J.~W., Icke, V., \& Dominik, C. 2001, \aap, 371, 205

\bibitem[{Skinner {et~al.}(1997)Skinner, Meixner, Barlow, Collison, Justtanont,
  Blanco, na, \& Ball}]{SMB_97_Egg}
Skinner, C.~J., Meixner, M., Barlow, M.~J., {et~al.} 1997, \aap, 328, 290

\bibitem[{Stanek {et~al.}(1995)Stanek, Knapp, Young, \&
  Phillips}]{SKY_95_molecular}
Stanek, K.~Z., Knapp, G.~R., Young, K., \& Phillips, T.~G. 1995, \apjs, 100,
  169

\bibitem[{Sylvester {et~al.}(1999)Sylvester, Kemper, Barlow, {de Jong}, Waters,
  Tielens, \& Omont}]{SKB_99_ohir}
Sylvester, R.~J., Kemper, F., Barlow, M.~J., {et~al.} 1999, \aap, 352, 587

\bibitem[{{van der Veen} \& Habing(1988)}]{VH_88_IRAScolors}
{van der Veen}, W. E. C.~J. \& Habing, H.~J. 1988, \aap, 194, 125

\bibitem[{{van Langevelde} {et~al.}(1990){van Langevelde}, {van der Heiden}, \&
  {van Schooneveld}}]{VVV_90_phaselags}
{van Langevelde}, H.~J., {van der Heiden}, R., \& {van Schooneveld}, C. 1990,
  \aap, 239, 193

\bibitem[{Vassiliadis \& Wood(1993)}]{VW_93_massloss}
Vassiliadis, E. \& Wood, P.~R. 1993, \apj, 413, 641

\bibitem[{Waters {et~al.}(1996)Waters, Molster, {de Jong}, Beintema, Waelkens,
  Boogert, Boxhoorn, de~Graauw, Drapatz, Feuchtgruber, Genzel, Helmich, Heras,
  Huygen, Izumiura, Justtanont, Kester, Kunze, Lahuis, Lamers, Leech, Loup,
  Lutz, Morris, Price, Roelfsema, Salama, Schaeidt, Tielens, Trams, Valentijn,
  Vandenbussche, {van den Ancker}, {van Dishoeck}, {van Winckel}, Wesselius, \&
  Young}]{WMJ_96_mineralogy}
Waters, L. B. F.~M., Molster, F.~J., {de Jong}, T., {et~al.} 1996, \aap, 315,
  L361

\bibitem[{Young(1995)}]{Y_95_CO3-2}
Young, K. 1995, \apj, 445, 872

\bibitem[{Yuasa {et~al.}(1999)Yuasa, Unno, \& Magono}]{YUM_99_distance}
Yuasa, M., Unno, W., \& Magono, S. 1999, \pasj, 51, 197

\bibitem[{Zubko \& Elitzur(2000)}]{ZE_00_WHya}
Zubko, V. \& Elitzur, M. 2000, \apjl, 544, L137

\end{thebibliography}

\appendix

\section{Observations --- Tables \& Figures}

\begin{table}
\caption{Programme stars. Distances are taken from 
$^{\rm{a}}$Hipparcos \citep{ESA_97_Hipparcos}, $^{\rm{b}}$\citet{VVV_90_phaselags}, 
$^{\rm{c}}$\citet{HBF_72_1612MHz}, $^{\rm{d}}$\citet{YUM_99_distance}, 
$^{\rm{e}}$\citet{HBP_86_luminosity}, $^{\rm{f}}$\citet{JHG_93_irc+10420}, 
$^{\rm{g}}$\citet{DGH_01_NMLCyg}}
\begin{center}
\begin{tabular}{l l l l}
\hline
\hline
object          & $\alpha$ (J2000) & $\delta$ (J2000) & $D$       \\
                &     &          & (kpc)\\
\hline
\object{T Cet}           & 00 21 46.27  & $-$20 03 28.9       & 0.238$^{\rm{a}}$\\
\object{WX Psc}          & 01 06 25.99  &   +12 35 53.4       & 0.74$^{\rm{b}}$\\
\object{OH 127.8+0.0}    & 01 33 51.19  &   +62 26 53.4       & 2.90$^{\rm{b}}$\\
\object{$o$ Cet}         & 02 19 20.793 & $-$02 58 39.51      & 0.128$^{\rm{a}}$\\ 
\object{IRC+50137}       & 05 11 19.37  &   +52 52 33.7       & 0.820$^{\rm{c}}$\\  
\object{$\alpha$ Ori}    & 05 55 10.305 &   +07 24 25.43      & 0.131$^{\rm{a}}$\\ 
\object{VY CMa}          & 07 22 58.33  & $-$25 46 03.2       & 0.562$^{\rm{a}}$\\
\object{$\alpha$ Sco}    & 16 29 24.461 & $-$26 25 55.21      & 0.185$^{\rm{a}}$\\
\object{V438 Oph}        & 17 14 39.78  &   +11 04 10.0       & \\
\object{AFGL~5379}       & 17 44 23.89  & $-$31 55 39.11      & 1.19$^{\rm{d}}$\\ 
\object{VX Sgr}          & 18 08 04.05  & $-$22 13 26.6       & 0.330$^{\rm{a}}$\\
\object{CRL~2199}        & 18 35 46.9   &   +05 35 48         & 2.48$^{\rm{d}}$\\
\object{OH 26.5+0.6}     & 18 37 32.52  & $-$05 23 59.4       & 1.37$^{\rm{b}}$\\
\object{OH 30.1$-$0.7}   & 18 48 41.5   & $-$02 50 29         & 1.77$^{\rm{e}}$\\
\object{OH 32.8$-$0.3}   & 18 52 22.19  & $-$00 14 13.9       & 5.02$^{\rm{b}}$\\ 
\object{OH 44.8$-$2.3}   & 19 21 36.56  &   +09 27 56.3       & 1.13$^{\rm{b}}$\\
\object{IRC+10420}       & 19 26 48.09  &   +11 21 16.7       & 5$^{\rm{f}}$\\
\object{NML Cyg}         & 20 46 25.7   &   +40 06 56         & 1.22$^{\rm{g}}$\\
\object{$\mu$ Cep}       & 21 43 30.461 &   +58 46 48.17      & 1.613$^{\rm{a}}$\\ 
\object{IRAS21554+6204} & 21 56 58.3   &   +62 18 43         & 2.03$^{\rm{d}}$\\ 
\object{OH 104.9+2.4}    & 22 19 27.9   &   +59 51 22         & 2.30$^{\rm{b}}$\\
\hline
\hline
\end{tabular}
\end{center}
\label{tab:obslist}
\end{table}

\begin{flushleft}
\begin{table}[!h]
\caption[]{Details of the observations. For each source the observed 
transitions are listed, together with the integrated observing time in 
seconds. The correction factor $f$ has been applied to our measurements, 
derived from standard measurements. The last column 
lists the observing dates.}
\begin{tabular}{l l l l l}
\hline
\hline
object          & transition         & $t_{\mathrm{int}}$ & $f$ &obs.~date \\ 
                &                    & (s)  & \\
\hline
\object{T Cet}           & CO(2$-$1) & 1800      & 1.07      & 03-Sep-02 \\
\object{WX Psc}          & CO(2$-$1) & 1800      & 0.90      & 22-Mar-01 \\
                         & CO(3$-$2) & 1200      & 1         & 02-Jul-00 \\
                         & CO(4$-$3) & 4800      & 1         & 21-Apr-00 \\
                         & CO(6$-$5) & 7320      & --        & 10-Oct-01 \\
                         & CO(7$-$6) & 7200      & --        & 09-Oct-01 \\
\object{OH 127.8+0.0}    & CO(2$-$1) & 1200      & 1.09      & 03-Sep-02 \\
                         & CO(3$-$2) & 2400      & 1         & 02-Jul-00 \\ 
                         & CO(4$-$3) & 5600      & 1         & 13-Apr-00 \\ 
\object{$o$ Cet}         & CO(2$-$1) & 600       & 1.08      & 03-Sep-02 \\
                         & CO(3$-$2) & 600       & 1         & 02-Jul-00 \\
\object{IRC+50137}       & CO(2$-$1) & 1800      & 0.96      & 06-Dec-00 \\
                         & CO(3$-$2) & 2400      & 1.10      & 05-Dec-00 \\
\object{$\alpha$ Ori}    & CO(3$-$2) & 5400      & 1         & 02-Jul-00 \\ 
\object{VY CMa}          & CO(2$-$1) & 3600      & 1         & 22-Mar-01 \\
                         & CO(3$-$2) & 2400      & 1.10      & 05-Dec-00 \\
                         & CO(6$-$5) & 8400      & --        & 10-Oct-01 \\
                         & CO(7$-$6) & 5400      & --        & 09-Oct-01 \\
\object{$\alpha$ Sco}    & CO(3$-$2) & 1800      & 1.10      & 04-Jul-00 \\
\object{V438 Oph}        & CO(2$-$1) & 1800      & 1.01      & 03-Sep-02 \\
\object{AFGL~5379}       & CO(3$-$2) & 1200      & 1         & 17-Apr-00 \\
\object{VX Sgr}          & CO(2$-$1) & 1860      & 1         & 22-Mar-01 \\
                         & CO(3$-$2) & 2400      & 1         & 18-Apr-00 \\
                         & CO(4$-$3) & 2400      & 1         & 04-Jul-00 \\
\object{CRL~2199}        & CO(2$-$1) & 1800      & 1         & 22-Mar-01 \\
                         & CO(3$-$2) & 1200      & 1         & 17-Apr-00 \\
                         & CO(4$-$3) & 8400      & 1         & 21-Apr-00 \\
\object{OH 26.5+0.6}     & CO(2$-$1) & 1800      & 0.95      & 22-Mar-01 \\
                         & CO(3$-$2) & 1200      & 1         & 17-Apr-00 \\
                         & CO(4$-$3) & 8400      & 1         & 21-Apr-00 \\
\object{OH 30.1$-$0.7}   & CO(3$-$2) & 1020      & 1         & 06-Jul-00 \\
\object{OH 32.8$-$0.3}   & CO(3$-$2) & 1200      & 1         & 18-Apr-00 \\
\object{OH 44.8$-$2.3}   & CO(3$-$2) & 2400      & 1         & 07-Jul-00 \\
\object{IRC+10420}       & CO(2$-$1) & 1800      & 0.95      & 22-Mar-01 \\
                         & CO(3$-$2) & 1200      & 1         & 17-Apr-00 \\
                         & CO(4$-$3) & 2400      & 1         & 21-Apr-00 \\
\object{NML Cyg}         & CO(2$-$1) & 1800      & 0.90      & 22-Mar-01 \\
                         & CO(3$-$2) & 1200      & 1         & 17-Apr-00 \\
                         & CO(4$-$3) & 1200      & 1         & 21-Apr-00 \\
                         & CO(6$-$5) & 4800      & --        & 10-Oct-01 \\
\object{$\mu$ Cep}       & CO(2$-$1) & 1800      & 1         & 20-Sep-02 \\
                         & CO(3$-$2) & 1200      & 1         & 18-Apr-00 \\
\object{IRAS 21554+6204} & CO(2$-$1) & 1920      & 1         & 20-Sep-02 \\
                         & CO(3$-$2) & 3600      & 1         & 07-Jul-00 \\
\object{OH 104.9+2.4}    & CO(2$-$1) & 2400      & 0.90      & 22-Mar-01 \\
                         & CO(3$-$2) & 2400      & 1         & 18-Apr-00 \\
                         & CO(4$-$3) & 4800      & 1.16      & 13-Apr-00 \\
                         &           & 2160      & 1         & 21-Apr-00 \\
                         & CO(6$-$5) & 11400     & --        & 10-Oct-01 \\
                         & CO(7$-$6) & 2400      & --        & 09-Oct-01 \\
\hline
\hline
\end{tabular}
\label{tab:obsdetails}
\end{table}
\end{flushleft}

\begin{flushleft}
\begin{table*}[!h]
\caption[]{Overview of observed line parameters. For each observed 
transition, the peak intensity ($T_{\mathrm{mb}}$) is measured, together 
with the r.m.s. values (Col.~4) and corresponding bin sizes
(Col.~5). Cols.~6 and 7 represent the velocity of the object 
($v_{\mathrm{LSR}}$) and the terminal velocity ($v_{\infty}$). The 
integrated line intensity (Col.~8) is determined by removing interstellar
absorption and emission from the profile, and integrating the remaining 
line profile. The accuracy on these
values is at least 10\% for the CO(2$-$1), (3$-$2) and 
(4$-$3) transitions and at least 30\% for the CO(6$-$5) and (7$-$6) 
transitions. In some cases the accuracy is 
deteriorated due to interstellar contamination.}
\begin{tabular}{l l l l l l l l}
\hline
\hline
object          & transition  & $T_{\mathrm{mb}}$ (K)& r.m.s. (K) & bin (MHz)& $V_{\mathrm{LSR}}$ (km s$^{-1}$) & $V_{\infty}$ (km s$^{-1}$) & $I$ (K km s$^{-1}$) \\
\hline
T Cet           & CO(2$-$1) & 0.44      & 0.056 & 0.3125 & +23.1 $\pm$ 0.5   & 6.7 $\pm$ 0.5  & 3.9 $\pm$ 0.4\\ 
WX Psc          & CO(2$-$1) & 2.35      & 0.034 & 0.6250 & +9.0 $\pm$ 0.5    & 20.2 $\pm$ 0.5 & 66 $\pm$ 7\\
                & CO(3$-$2) & 2.91      & 0.043 & 1.2500 & +9.2 $\pm$ 0.5    & 20.3 $\pm$ 0.5 & 82 $\pm$ 8\\
                & CO(4$-$3) & 1.86      & 0.090 & 0.9375 & +9.0 $\pm$ 0.5    & 20.6 $\pm$ 0.5 & 50 $\pm$ 5\\
                & CO(6$-$5) & 0.45      & 0.224 & 2.5000 & +8.4 $\pm$ 1.5    & 17 $\pm$ 3     & 10 $\pm$ 5\\
                & CO(7$-$6) & 0.82      & 0.378 & 3.1250 & +9.8 $\pm$ 1.5    & 21 $\pm$ 3     & 23 $\pm$ 11\\
OH 127.8+0.0    & CO(2$-$1) & 0.28      & 0.048 & 0.6250 & $-$56 $\pm$ 3     & 13 $\pm$ 2     & 5.5 $\pm$ 1.6\\ 
                & CO(3$-$2) & 0.68      & 0.050 & 0.6250 & $-$55 $\pm$ 3     & 13 $\pm$ 2     & 12 $\pm$ 1\\ 
                & CO(4$-$3) & 0.15      & 0.057 & 1.5625 & complex           & complex        & 23 $\pm$ 2\\ 
$o$ Cet         & CO(2$-$1) & 13.80     & 0.170 & 0.1562 & +46.3 $\pm$ 0.1   & 8 $\pm$ 1      & 64 $\pm$ 6\\ 
                & CO(3$-$2) & 21.61     & 0.134 & 0.3125 & +46.4 $\pm$ 0.1   & 8 $\pm$ 2      & 108 $\pm$ 11\\
IRC+50137       & CO(2$-$1) & 1.37      & 0.033 & 0.4688 & +2.8 $\pm$ 0.5    & 19.1 $\pm$ 0.5 & 37 $\pm$ 4\\
                & CO(3$-$2) & 1.44      & 0.047 & 0.9375 & +3.2 $\pm$ 0.5    & 18.5 $\pm$ 0.5 & 39 $\pm$ 4\\
$\alpha$ Ori    & CO(3$-$2) & complex   & 0.043 & 0.3125 & +3.4 $\pm$ 0.5    & 15.7 $\pm$ 0.5 & 50 $\pm$ 5\\ 
VY CMa          & CO(2$-$1) & complex   & 0.033 & 0.6250 & +25 $\pm$ 3       & 47 $\pm$ 3     & 66 $\pm$ 7\\
                & CO(3$-$2) & 3.00      & 0.043 & 1.2500 & +25 $\pm$ 3       & 47 $\pm$ 3     & 173 $\pm$ 17\\
                & CO(6$-$5) & 4.37      & 0.469 & 3.7500 & +27 $\pm$ 2       & 48 $\pm$ 3     & 257 $\pm$ 77\\
                & CO(7$-$6) & 7.41      & 0.908 & 3.7500 & +29 $\pm$ 2       & 44 $\pm$ 3     & 433 $\pm$ 130\\
$\alpha$ Sco    & CO(3$-$2) & --        & 0.036 & 1.8750 & --                & --             & --\\
V438 Oph        & CO(2$-$1) & 0.24      & 0.058 & 0.3125 & +9.7 $\pm$ 0.5    & 4.3 $\pm$ 1.0  & 1.1 $\pm$ 0.2\\ 
AFGL~5379       & CO(3$-$2) & 2.76      & 0.056 & 1.2500 & $-$22.7 $\pm$ 1.5 & 24 $\pm$ 2     & 84 $\pm$ 8\\
VX Sgr          & CO(2$-$1) & complex   & 0.048 & 0.4688 & +6.4 $\pm$ 1.0    & 25 $\pm$ 1     & 31 $\pm$ 3\\
                & CO(3$-$2) & 2.37      & 0.059 & 0.6250 & +6.9 $\pm$ 1.0    & 26 $\pm$ 1     & 97 $\pm$ 10\\
                & CO(4$-$3) & 1.22      & 0.237 & 1.2500 & +6.4 $\pm$ 0.5    & 22 $\pm$ 2     & 42 $\pm$ 4\\
CRL~2199        & CO(2$-$1) & 1.16      & 0.033 & 0.6250 & +33.7 $\pm$ 0.2   & 17.6 $\pm$ 0.5 & 26 $\pm$ 3\\
                & CO(3$-$2) & 1.25      & 0.034 & 1.2500 & +33.5 $\pm$ 0.2   & 17.9 $\pm$ 0.5 & 28 $\pm$ 3\\
                & CO(4$-$3) & 0.98      & 0.066 & 1.2500 & +33.2 $\pm$ 0.5   & 18 $\pm$ 1     & 22 $\pm$ 2\\
OH 26.5+0.6     & CO(2$-$1) & contamin. & 0.049 & 0.3125 & contamin.         & contamin.      & 9 $\pm$ 3\\
                & CO(3$-$2) & 1.05      & 0.046 & 0.9375 & +26.9 $\pm$ 0.5   & 18 $\pm$ 1     & 23 $\pm$ 2\\
                & CO(4$-$3) & 0.86      & 0.051 & 0.9375 & +27.7 $\pm$ 0.3   & 15.9 $\pm$ 0.5 & 19 $\pm$ 2\\
OH 30.1$-$0.7   & CO(3$-$2) & contamin. & 0.076 & 0.9375 & contamin.         & contamin.      & contamin.\\
OH 32.8$-$0.3   & CO(3$-$2) & contamin. & 0.047 & 0.9375 & contamin.         & contamin.      & contamin.\\
OH 44.8$-$2.3   & CO(3$-$2) & 0.63      & 0.047 & 0.9375 & $-$70.3 $\pm$ 0.2 & 17.7 $\pm$ 0.5 & 15 $\pm$ 2\\
IRC+10420       & CO(2$-$1) & 1.65      & 0.034 & 0.4688 & +75 $\pm$ 1       & 43 $\pm$ 3     & 95 $\pm$ 10\\
                & CO(3$-$2) & 3.23      & 0.075 & 0.9375 & +75 $\pm$ 1       & 45 $\pm$ 3     & 180 $\pm$ 18\\
                & CO(4$-$3) & 2.84      & 0.078 & 1.8750 & +76.1 $\pm$ 0.5   & 42 $\pm$ 2     & 150 $\pm$ 15\\
NML Cyg         & CO(2$-$1) & complex   & 0.039 & 0.3125 & $-$2 $\pm$ 3      & 33 $\pm$ 3     & 99 $\pm$ 10\\
                & CO(3$-$2) & complex   & 0.104 & 0.6250 & $-$2 $\pm$ 3      & 33 $\pm$ 3     & 210 $\pm$ 21\\
                & CO(4$-$3) & complex   & 0.106 & 1.2500 & $-$1 $\pm$ 2      & 34 $\pm$ 2     & 133 $\pm$ 13\\
                & CO(6$-$5) & complex   & 0.232 & 2.5000 & +2 $\pm$ 2        & 34 $\pm$ 2     & 111 $\pm$ 56\\
$\mu$ Cep       & CO(2$-$1) & complex   & 0.025 & 1.2500 & +22 $\pm$ 2       & 33 $\pm$ 3     & 2.5 $\pm$ 0.3\\ 
                & CO(3$-$2) & complex   & 0.066 & 0.9375 & +21 $\pm$ 3       & 35 $\pm$ 3     & 14 $\pm$ 3\\
IRAS 21554+6204 & CO(2$-$1) & 0.58      & 0.041 & 0.4688 & $-$19 $\pm$ 1     & 19 $\pm$ 2     & 12 $\pm$ 1\\ 
                & CO(3$-$2) & 0.49      & 0.036 & 0.9375 & $-$19.0 $\pm$ 0.5 & 18.1 $\pm$ 0.5 & 10 $\pm$ 1\\
OH 104.9+2.4    & CO(2$-$1) & 0.21      & 0.026 & 0.6250 & $-$26 $\pm$ 1     & 18.6 $\pm$ 0.5 & 5.4 $\pm$ 0.5\\
                & CO(3$-$2) & 0.43      & 0.043 & 0.9375 & $-$25 $\pm$ 1     & 18.3 $\pm$ 0.5 & 11 $\pm$ 1\\
                & CO(4$-$3) & 0.11      & 0.042 & 1.8455 & $-$26 $\pm$ 1     & 18 $\pm$ 1     & 3.6 $\pm$ 0.5\\
                & CO(6$-$5) & --        & 0.224 & 3.7500 & --                & --             & --\\
                & CO(7$-$6) & --        & 1.011 & 3.1250 & --                & --             & --\\
\hline
\hline
\end{tabular}
\label{tab:transitions}
\end{table*}
\end{flushleft}

\clearpage

\begin{table}
\caption{Observed CO line transitions in semi-regular variables and AGB stars, obtained from the literature. The listed stars are observed in CO(4$-$3), CO(3$-$2) and CO(2$-$1). The references are $^{\mathrm{a}}$\citet{BKO_00_rvboo}, $^{\mathrm{b}}$\citet{KO_99_COcatalogue}, $^{\mathrm{c}}$\citet{KYL_98_COsurvey}, $^{\mathrm{d}}$\citet{Y_95_CO3-2}, $^{\mathrm{e}}$\citet{SKY_95_molecular}.}
\begin{center}
\begin{tabular}{l l l l l}
\hline
\hline
source          & line      & $I$(K km s$^{-1}$) & telescope & ref. \\
\hline
\object{RV Boo} & CO(4$-$3) & 8.87               & JCMT      & a\\
                & CO(3$-$2) & 6.81               & JCMT      & a\\
                & CO(2$-$1) & 4.47               & JCMT      & a\\
\object{X Her}  & CO(4$-$3) & 42.17              & JCMT      & b\\
                & CO(3$-$2) & 25.69              & JCMT      & b\\
                & CO(3$-$2) & 13.3 $\pm$ 1.3     & CSO       & c\\
                & CO(2$-$1) & 6.4 $\pm$ 0.4      & CSO       & c\\
\object{R LMi}  & CO(4$-$3) & 7.1                & CSO       & d\\
                & CO(3$-$2) & 9.5                & CSO       & d\\
                & CO(2$-$1) & 2.72 $\pm$ 0.38    & CSO       & c\\
\object{R Hya}  & CO(4$-$3) & 46.1               & CSO       & d\\
                & CO(3$-$2) & 37                 & CSO       & d\\
                & CO(3$-$2) & 22.2 $\pm$ 2.2     & CSO       & c\\
                & CO(2$-$1) & 4.9 $\pm$ 0.5      & CSO       & c\\
\object{S Vir}  & CO(4$-$3) & 2.3                & CSO       & d\\
                & CO(3$-$2) & 2.5                & CSO       & d\\
                & CO(2$-$1) & 0.64 $\pm$ 0.25    & CSO       & c\\
\object{S CrB}  & CO(4$-$3) & 4.9                & CSO       & d\\
                & CO(3$-$2) & 9.1                & CSO       & d\\
                & CO(2$-$1) & 2.53 $\pm$ 0.51    & CSO       & c\\
\object{RU Her} & CO(4$-$3) & 5.7                & CSO       & d\\
                & CO(3$-$2) & 9.4                & CSO       & d\\
                & CO(2$-$1) & 2.3 $\pm$ 0.2      & CSO       & c\\
\object{$\chi$ Cyg}  & CO(4$-$3) & 52.7          & CSO       & d\\
                & CO(3$-$2) & 63                 & CSO       & d\\
                & CO(3$-$2) & 41.5               & CSO       & e\\
                & CO(2$-$1) & 28.8 $\pm$ 0.7     & CSO       & c\\
\hline
\hline
\end{tabular}
\label{tab:litvalues}
\end{center}
\end{table}

\begin{figure}
  \resizebox{\hsize}{!}{\includegraphics{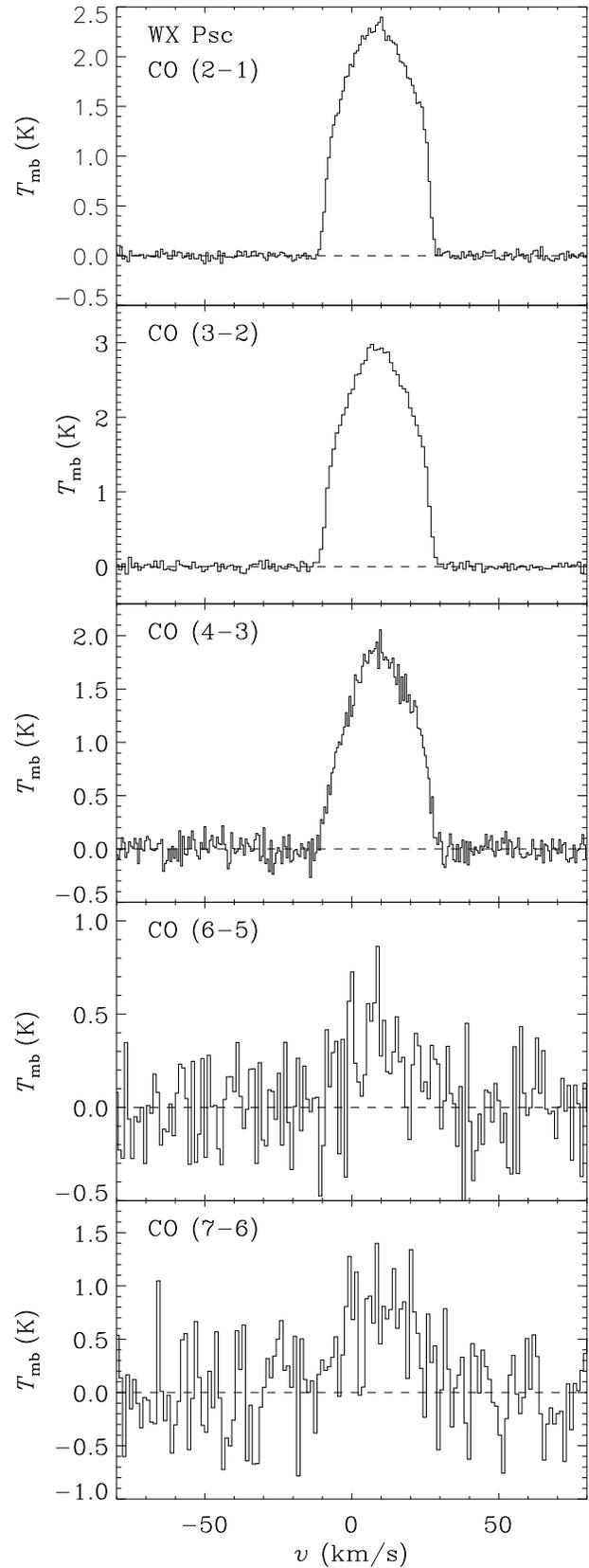}}
  \caption{JCMT observations of rotational transitions of CO observed in 
WX Psc. See text for details on data reduction and analysis.}
  \label{fig:wxpsc}
\end{figure}

\begin{figure}
  \resizebox{\hsize}{!}{\includegraphics{H4224F12.eps}}
  \caption{Idem -- T Cet}
  \label{fig:tcet}
\end{figure}

\begin{figure}
  \resizebox{\hsize}{!}{\includegraphics{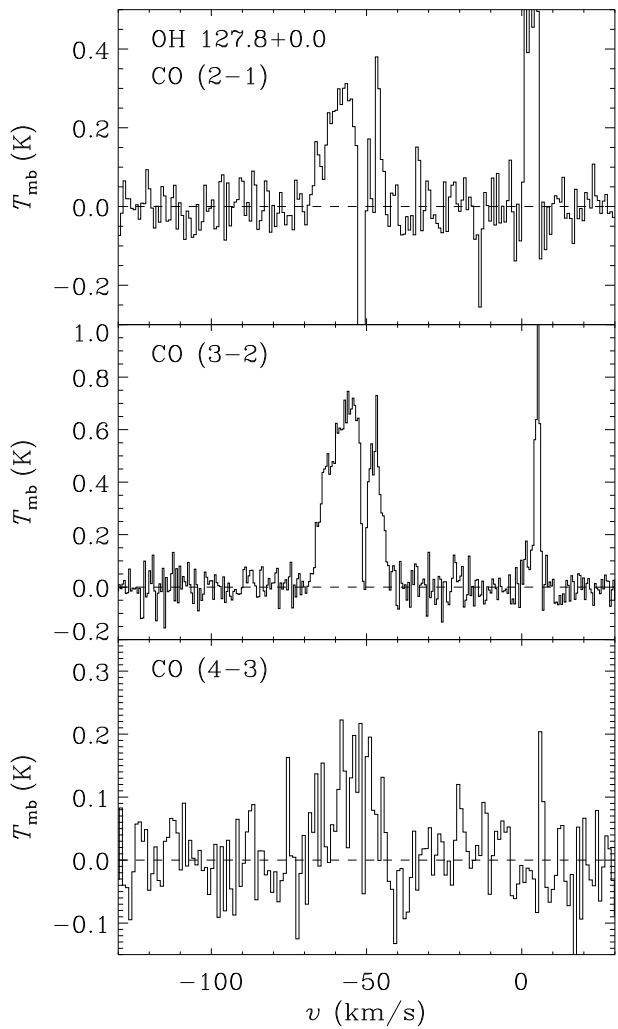}}
  \caption{Idem -- OH 127.8+0.0}
  \label{fig:oh127}
\end{figure}

\begin{figure}
  \resizebox{\hsize}{!}{\includegraphics{H4224F14.eps}}
  \caption{Idem -- $o$ Cet}
  \label{fig:ocet}
\end{figure}

\begin{figure}
  \resizebox{\hsize}{!}{\includegraphics{H4224F15.eps}}
  \caption{Idem -- IRC +50137}
  \label{fig:irc50}
\end{figure}

\begin{figure}
  \resizebox{\hsize}{!}{\includegraphics{H4224F16.eps}}
  \caption{Idem -- $\alpha$ Ori}
  \label{fig:aori}
\end{figure}

\begin{figure}
  \resizebox{\hsize}{!}{\includegraphics{H4224F17.eps}}
  \caption{Idem -- $\alpha$ Sco}
  \label{fig:asco}
\end{figure}

\begin{figure}
  \resizebox{\hsize}{!}{\includegraphics{H4224F18.eps}}
  \caption{Idem -- V438 Oph}
  \label{fig:v438oph}
\end{figure}

\begin{figure}
  \resizebox{\hsize}{!}{\includegraphics{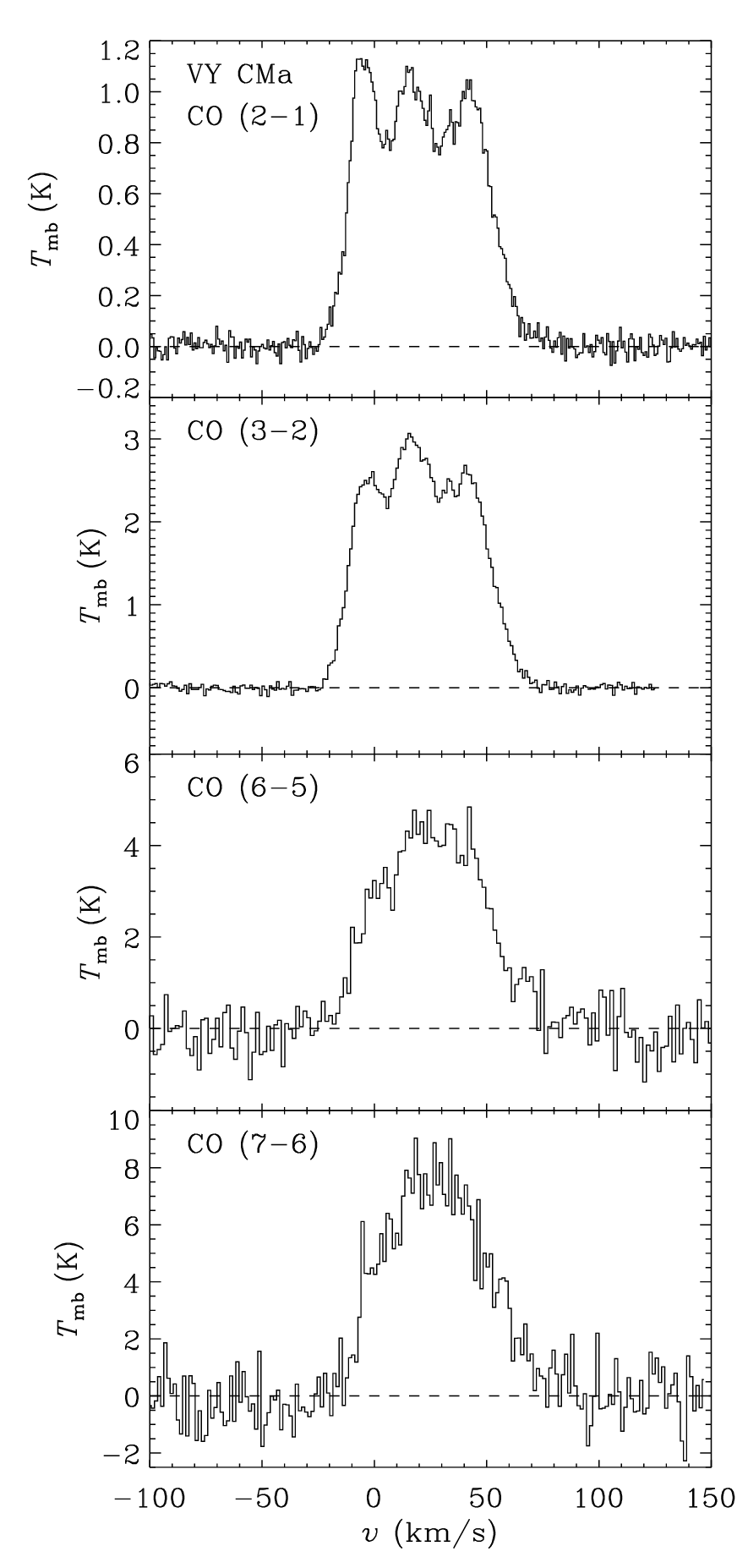}}
  \caption{Idem -- VY CMa}
  \label{fig:vycma}
\end{figure}

\begin{figure}
  \resizebox{\hsize}{!}{\includegraphics{H4224F20.eps}}
  \caption{Idem -- AFGL 5379}
  \label{fig:gl}
\end{figure}

\begin{figure}
  \resizebox{\hsize}{!}{\includegraphics{H4224F21.eps}}
  \caption{Idem -- VX Sgr}
  \label{fig:vxsgr}
\end{figure}

\begin{figure}
  \resizebox{\hsize}{!}{\includegraphics{H4224F22.eps}}
  \caption{Idem -- CRL 2199}
  \label{fig:crl}
\end{figure}

\begin{figure}
  \resizebox{\hsize}{!}{\includegraphics{H4224F23.eps}}
  \caption{Idem -- OH 30.1$-$0.7}
  \label{fig:oh30}
\end{figure}

\clearpage
\begin{figure}
  \resizebox{\hsize}{!}{\includegraphics{H4224F24.eps}}
  \caption{Idem -- OH 26.5+0.6}
  \label{fig:oh26}
\end{figure}

\begin{figure}
  \resizebox{\hsize}{!}{\includegraphics{H4224F25.eps}}
  \caption{Idem -- OH 32.8$-$0.3}
  \label{fig:oh32}
\end{figure}

\begin{figure}
  \resizebox{\hsize}{!}{\includegraphics{H4224F26.eps}}
  \caption{Idem -- OH 44.8$-$2.3}
  \label{fig:oh44}
\end{figure}

\begin{figure}
  \resizebox{\hsize}{!}{\includegraphics{H4224F27.eps}}
  \caption{Idem -- IRC +10420}
  \label{fig:irc10}
\end{figure}

\begin{figure}
  \resizebox{\hsize}{!}{\includegraphics{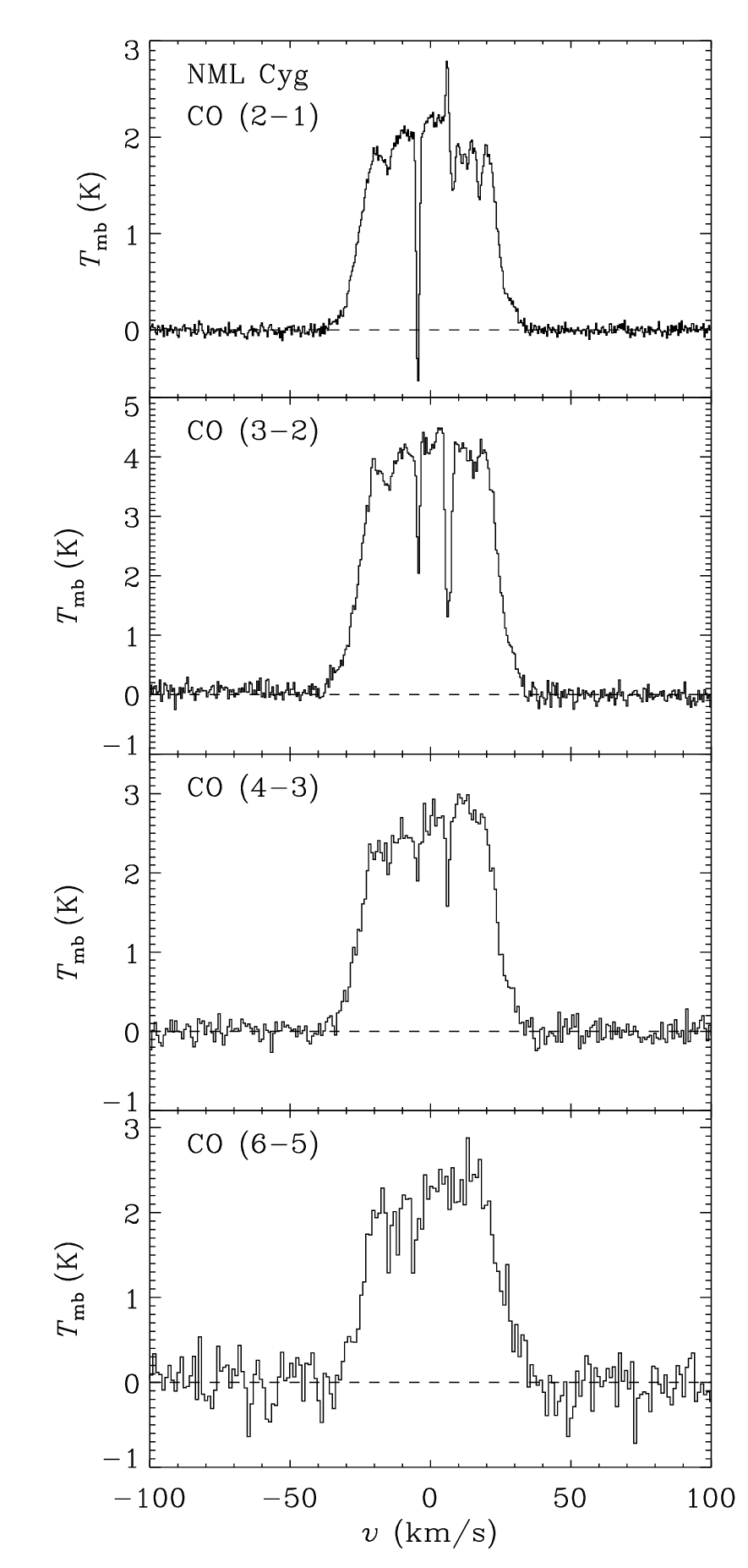}}
  \caption{Idem -- NML Cyg}
  \label{fig:nmlcyg}
\end{figure}

\begin{figure}
  \resizebox{\hsize}{!}{\includegraphics{H4224F29.eps}}
  \caption{Idem -- $\mu$ Cep}
  \label{fig:mucep}
\end{figure}

\begin{figure}
  \resizebox{\hsize}{!}{\includegraphics{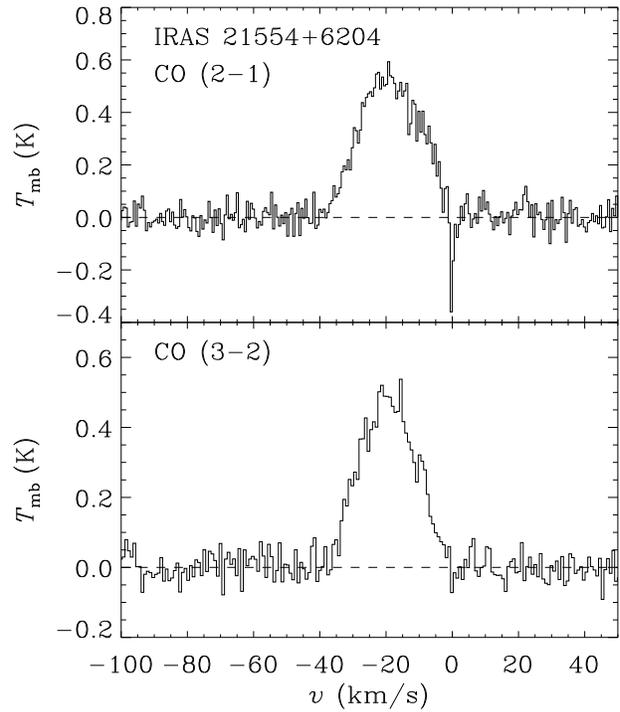}}
  \caption{Idem -- IRAS 21554+6204}
  \label{fig:iras21}
\end{figure}

\begin{figure}
  \resizebox{\hsize}{!}{\includegraphics{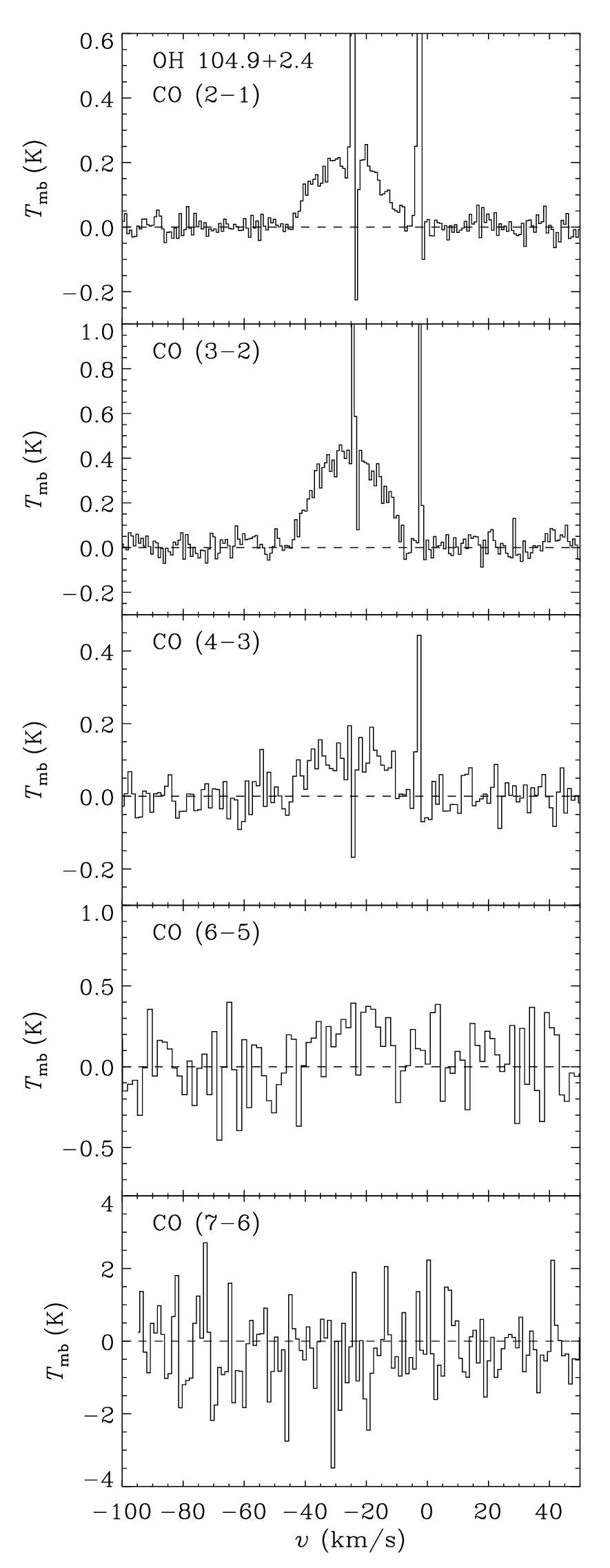}}
  \caption{Idem -- OH 104.9+2.4}
  \label{fig:oh104}
\end{figure}

\end{document}